\documentclass[useAMS]{mn2e}
\usepackage{times}
\usepackage{amssymb}
\usepackage{rotating}
\input{epsf}

\title[Physical PSD model]
{A physical model for the continuum variability and QPO 
in accreting black holes}
\author[A. Ingram \& Chris Done]
{Adam
Ingram$^{1}\thanks{E-mail:a.r.ingram@durham.ac.uk}$ \&
Chris Done$^{1}$\\
$^1$Department of Physics, University of Durham, South Road,
Durham DH1 3LE, UK\\
}

\date{Submitted to MNRAS}
\pagerange{\pageref{firstpage}--\pageref{lastpage}} \pubyear{2010}

\begin{document}

\topmargin = -0.5cm

\maketitle

\label{firstpage}

\begin{abstract}

The power spectra of black hole binaries have been well studied
for decades, giving a very detailed phenomenological picture of the
variability properties and their correlation with
the energy spectrum (spectral state) of the source.
Here we take the
truncated disc/hot inner flow picture which can describe the spectral
changes, and show that propagating mass accretion rate fluctuations in
the hot flow can match the broad band power spectral properties
seen in black hole binaries, i.e. give approximately
band limited noise between a low and high frequency break. 
The low frequency break marks the viscous timescale at the outer
edge of the hot inner flow, which is the inner edge of the truncated
disc. The fluctuations in mass accretion rate propagate towards the
central object in a finite time meaning the high frequency break is more
complex than simply the viscous timescale at the inner edge of the hot
flow because fluctuations on timescales shorter than the propagation
time are incoherent. The model also predicts the
Lense-Thirring precession timescale of the hot flow, as this is set by the
combination of inner and outer radius of the flow, together with its
surface density which is self consistently calculated from the propagating
fluctuations. We show that this naturally gives the observed relation
between the low frequency break and QPO frequency as the
outer radius of the flow moves inwards, and that this model
predicts many of the observed QPO properties such as correlation of
coherence with frequency, and of the recently discovered
correlation of frequency with flux on short timescales.

We fit this total model of the variability to a sequence of 5 observed
power spectra from the bright black hole binary XTE J1550-564 as the
source transitioned from a low/hard to very high state.  This is the
first time that a power spectrum from a black hole binary has been fit
with a physical model for the variability. The data are well fit if the
inner radius of the flow remains constant, while the outer radius sweeps
inwards from $\sim 75-12R_g$. This range of radii is the same range as
required by models of the energy spectral evolution, giving the first
self consistent description of the evolution of both the spectrum and
variability of BHB.

\end{abstract}


%==============================================
\section{Introduction} \label{sec:introduction}
%==============================================

Emission observed from Black Hole Binaries (BHBs) is variable on both
long and short timescales. On the longest timescales, changes in mass
accretion rate drive changes in spectral state. At the lowest
luminosities, the system is typically seen in the low/hard state,
where the spectrum is dominated by a hard (photon index $\Gamma<2$)
power law tail, peaking at 100~keV. As the source brightens, there is
an increasing contribution from the disc at low energies and the tail
softens towards $\Gamma\sim 2$ (intermediate state). The tail can then
either remain strong as the source flux increases, giving the very
high state ($\Gamma\sim 2.5$) or it can carry only a very small
fraction of the luminosity, giving the high/soft state ($\Gamma \sim
2.2$) (e.g. see reviews by Remillard \& McClintock 2006; Done,
Gierlinski \& Kubota 2007, hereafter DGK07).

Much of this long term spectral evolution can be successfully
described by the truncated disc model (DGK07). Here, the assumption is
that the cool, geometrically thin, optically thick disc is truncated
at some radius, $r_o$, which is much larger than the last stable
orbit, $r_{lso}$ at the lowest luminosities.  The inner accretion flow
extends from $r_o$ to an innermost radius, $r_{i}$, forming some
sort of hot, geometrically thick, optically thin (optical depth
$\tau\sim 1.5$) flow, similar to an Advection Dominated Accretion Flow
(ADAF, Narayan \& Yi 1995).  Decreasing the truncation radius of the
thin disc with increasing luminosity (and therefore average mass
accretion rate) leads to a stronger disc component and greater
illumination of the hot flow by the cool disc photons and hence to
a softer Comptonised tail. This spectral evolution has a natural end
when $r_o\sim r_{lso}$, marking the transition to the disc dominated
soft states seen at higher luminosities.

There is also rapid variability on timescales of 100-0.1s
which is almost exclusively linked to the Comptonised tail (e.g. Churazov
et al 2001). This is as expected as the disc emission can only vary on
a viscous timescale of $t_{visc}=4.5\times 10^{-3} \alpha^{-1}
(h/r)^{-2} (R/6R_g)^{3/2} (M/10M_{\sun})$~s i.e. 500~s at the last stable
orbit for fiducial thin disc parameters of $h/r\sim 0.01$ and
$\alpha=0.1$.

A power spectral analysis of the rapid variability shows that it is
generally composed of a broadband noise continuum, with a low
frequency quasi-periodic oscillation (LF QPO) superimposed on
this. The broadband noise can be very roughly characterized as a twice
broken power law, with $fP_f\propto f~$ at the lowest frequencies,
breaking to $fP_f \propto f^0$ above the low frequency break $f_b$,
and then breaking again to $fP_f \propto f^{-1}$ above a high
frequency break at $f_h$. The low frequency break is tightly
correlated with the spectral state of the source, with $f_b\sim
0.01$~Hz for the dimmest low/hard states, and increasing to $\sim
1$~Hz as the source softens through the bright low/hard states into
the intermediate state. The LF QPO properties are also tightly
correlated with this power spectral and energy spectral evolution,
with QPO frequency $f_{QPO}\sim 10f_b$ increasing from 0.1-10~Hz, as
the coherence, power and harmonic content of this feature also
increase (e.g. van der Klis et al 2004). Again, this can be
interpreted qualitatively in the truncated disc picture if the low
frequency break and QPO are somehow associated with the truncation
radius. As the disc extends further inwards, the spectrum softens, and
all frequencies associated with this radius increase (e.g. DGK07).

However, to go beyond this into a quantitative description requires a
specific model for both the broadband noise and QPO. This is
problematic, despite both having been known for many decades (e.g. van
der Klis 1989). There are multiple potential models for the LF QPO in
the literature which fall into 2 main categories: those associated
with a geometrical misalignment of the accretion flow and black hole
spin (Stella \& Vietri 1998; Fragile, Mathews \& Wilson 2001;
Schnittman 2005; Schnittman et al 2006; Ingram, Done \& Fragile 2009,
hereafter IDF09), and those associated with wave modes of the
accretion flow (Wagoner et al 2001; Titarchuk \& Oscherovich 1999;
Cabanac et al 2008).  Most of these concentrate on matching the QPO
frequency, but the spectrum of the QPO gives additional
constraints. This is similar to that of the spectrum of the broadband
variability, showing that they both arise predominantly from the
Comptonising region rather than the disc (e.g Gilfanov et al 2003; 
Sobolewska \& Zycki
2006), favoring models in which the modulation arises
directly from the Comptonised emission e.g. IDF09, where the QPO is set
by Lense-Thirring (vertical) precession of the entire hot inner flow
interior to the disc truncation radius at $r_o$, or by a mode of the
hot inner flow (Cabernac et al 2008). 

The physical origin for the viscosity of the flow is the
Magneto-Rotational Instability (MRI: Balbus \& Hawley 1991).  This is
inherently variable, with large fluctuations in all quantities, both
spatially and temporally (Krolik \& Hawley 2002), making it a natural
origin for the broad band noise (Noble \& Krolik 2009; Ingram \& Done
2010; Dexter \& Fragile 2011). However, these fluctuations also
effectively shred any coherent wave modes in the flow (see
e.g. Reynolds \& Miller 2009; Henisey et al 2009). This effectively
rules out trapped wave propagation as the origin of the LF QPO,
leaving Lense-Thirring precession as the most likely
candidate. Precession of the entire hot flow from $r_o$ to $r_{i}$
can match the observed LF QPO frequency in both BHBs (IDF09)
\textit{and} neutron stars (NS; Ingram \& Done 2010), and provides a
clear mechanism to match the spectrum as this is a modulation of the
Comptonising region.

Thus the entire power spectrum can be explained by MRI fluctuations in
a hot flow, which is also precessing around the black hole.
However, the power spectrum does not represent all the information
contained in the variability, as it uses only Fourier amplitudes, not
phases. This is important as the light curves contain additional
correlations which give a linear rms-flux relation (Uttley \& McHardy
2001). Short segments of a longer light curve have a mean, $I$, and
variance, $\sigma^2$, which are related such that $I\propto \sigma$
(after binning: Uttley \& McHardy 2001). This is equivalent to the
flux on these timescales having a log-normal distribution (Negoro et
al 2000), and rules out simple models of the variability where the
light curve is made from adding together multiple, uncorrelated events
(Uttley \& Mchardy 2001; see also DGK07). Instead, this can be
produced if the light curve is made from a multiplicative process,
rather than an additive one. 
Again, the MRI in the hot flow gives a
physical interpretation to this. The MRI at large radii produces
intrinsic fluctuations in the density of the flow.  These fluctuations
propagate down to smaller radii on a viscous timescale, so all higher
frequency fluctuations are smoothed out. These smoothed and lagged
fluctuations in mass accretion rate modulate the MRI fluctuations produced
by the next radius, and so on, down to the smallest radii in the flow
(Luybarski 1997: hereafter L97). This naturally produces a light curve which
has an rms-flux relation/log-normal flux distribution (Kotov et al 2001:
hereafter K01; Arevalo \& Uttley 2006: hereafter AU06; also see Misra \&
Zdziarski 2008).

Here we use these ideas to build a model for the entire power
spectrum, where the broadband noise arises from propagation of MRI
fluctuations through the hot flow from $r_o$ to $r_{i}$ and the LF
QPO arises from Lense-Thirring precession {\em of the same hot flow}.
We fit this model to a sequence of power spectra of XTE J1550-564
observed during a transition from low/hard to intermediate state, and
show that these give a good {\em quantitative} fit for $r_o$ moving
from $\sim 75-10 R_g$, as also required to explain the sequence in energy
spectra during this transition. This gives very strong support to the
underlying truncated disc geometry, and for Lense-Thirring precession
as the physical origin for the LF QPO. 

\begin{figure}
\centering
\leavevmode  \epsfxsize=6.5cm \epsfbox{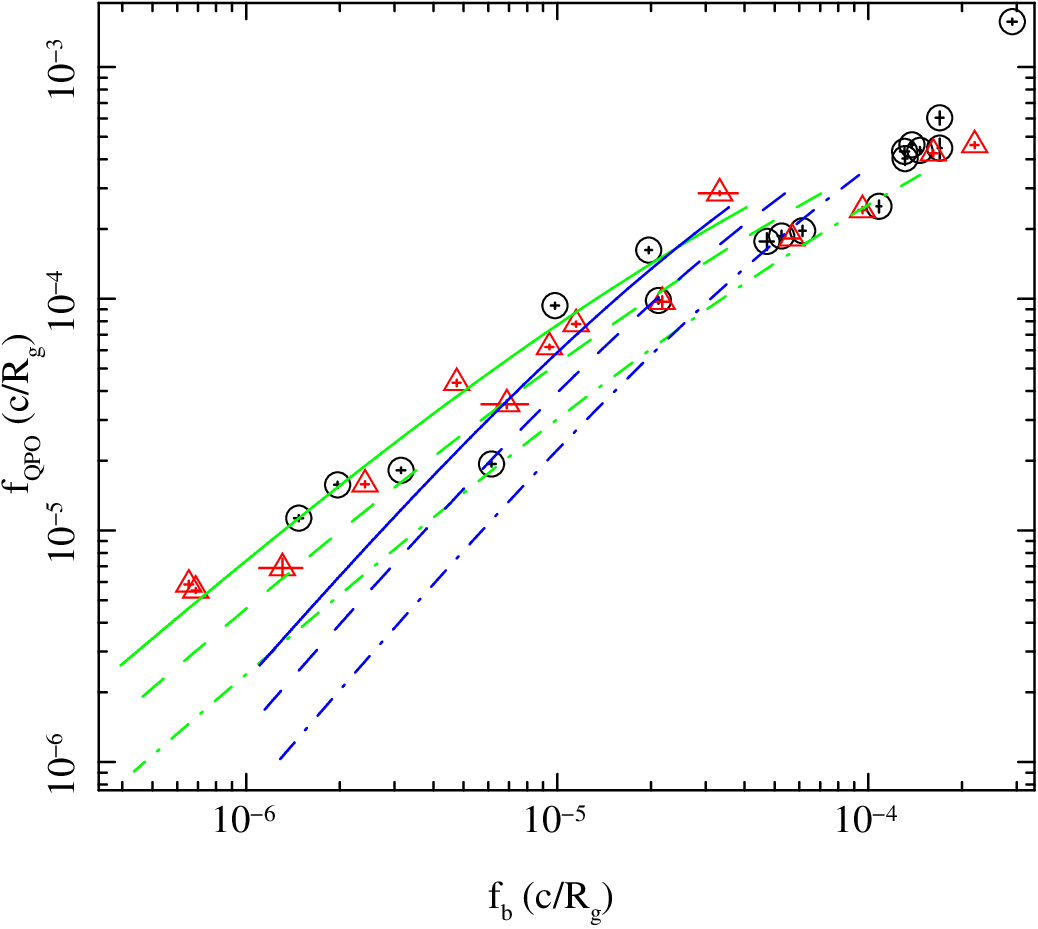}
\caption{
The QPO-break relation plotted in dimensions of $c/R_g$ for a
fiducial mass of $10$ and $1.4M_{\sun}$ for black holes
(black circles) and neutron stars (red triangles) respectively.
The fact that these frequencies lie on the same relationship
for the two objects implies a common physical origin.
The lines are predictions of the simplified model for spins
of $a_*=0.2$ (dot-dashed), $0.5$ (dashed) and $0.998$ (solid) with
$r_o$ ranging from $r_i$ to $100$ and $r_i=r_{bw}$. For the
blue lines, we assume the viscous frequency to be proportional
to the Keplerian frequency. For the green lines, we assume
$f_{visc}=Br^{-m}f_k$ where $B=0.03$ and $m=0.5$ and find
good agreement with the trend in the data.
}
\label{fig:wkplot}
\end{figure}

%====================================
\section{The simplified model} 
\label{sec:mod}
%====================================

We first introduce the simplified toy model considered in
IDF09 and Ingram \& Done (2010), whereby the low frequency break,
$f_b$, occurs at the viscous frequency of the truncation radius and
the QPO frequency, $f_{QPO}$, is the precession frequency of the
flow. Using the Shakura Sunyaev (1973) viscosity prescription, 
$f_b=f_{visc}(r_o)=\alpha (h/r)^2 f_k(r_o)= -v_r(r_o)/R_o$ where $\alpha$
is the viscosity parameter, $h/r$ the flow semi-thickness, $f_k$ the
Keplerian frequency and $v_r$ is the infall velocity.  The
Lense-Thirring precession frequency is then a weighted average of the
point particle precession frequency at each radius in the hot flow, so 
\begin{equation}
f_{prec} = f_{QPO} = \frac{\int_{r_{i}}^{r_{o}} f_{LT}f_{k}\Sigma r^3 dr}
{\int_{r_{i}}^{r_{o}} f_{k}\Sigma r^3 dr}
\label{eqn:fprec}
\end{equation}
(Liu \& Melia 2002) where $r_{i}$ is the innermost point of the flow
(i.e. the surface density is negligible interior to this), $\Sigma$ is the
surface density and
\begin{equation}
f_{LT}=f_k \left[1-\sqrt{1-\frac{4a_*}{r^{3/2}}+\frac{3a_*^2}{r^2}}\right]
\end{equation}
is the point particle Lense-Thirring precession frequency
for a dimensionless spin parameter $a_*$
(Merloni et al 1999). Here, $r$ is dimensionless, expressed in units of
$R_g=GM/c^2$. Solving this assuming a  power law form for
surface density in the hot flow 
$\Sigma \propto r^{-\zeta}$ between an inner and outer radius for the
hot flow $r_i$ and $r_o$ (Fragile et al 2007, IDF09) gives
\begin{equation}
f_{QPO}=\frac{(5-2\zeta)}{\pi(1+2\zeta)}
\frac{a_*[1-(r_i/r_o)^{1/2+\zeta}]}
{r_o^{5/2-\zeta}r_i^{1/2+\zeta}
[1-(r_i/r_o)^{5/2-\zeta}]}\frac{c}{R_g}.
\end{equation}

Hence this model predicts the relation between $f_{QPO}$ and $f_b$
which can be compared to the multiple observations of these
frequencies in both black holes and neutron stars (e.g. Wijnands \&
van der Klis 1999; Klein-Wolt \& van der Klis 2008). The observed relation
is continuous, implying that these frequencies show the same behavior
in both sources, i.e. that neither can depend strongly on any property
of the neutron star surface but are instead set by the accretion flow
itself. We re-plot this data in Figure \ref{fig:wkplot}, normalizing the
frequencies by mass for a fiducial mass of $1.4$ and $10M_\odot$ for
neutron stars (red triangles) and black holes (black circles),
respectively. This shows even more clearly that the two different types
of object show the same observed relation between these frequencies 
as they now occupy the same range. 

The blue lines show the prediction of the toy model, where the hot
flow has constant $\alpha=0.2$ and $h/r=0.2$, surface density constant
with radius (i.e. $\zeta=0$: Fragile et al 2007) between $r_o$ and
$r_i$, where $r_i$ is given by the bending wave radius. Warps in a large
scale height flow are communicated via bending waves which have wavelength
$\lambda \propto r^{9/4}$ and so are smooth at large $r$ and oscillatory
at small $r$. The bending wave radius ($r_i=3.0(h/r)^{-4/5}a_*^{2/5}$;
Fragile et al 2007; 2009; Fragile 2009; IDF09) marks the transition
between the two regimes. We show $a_*=0.2$ (dot-dashed), $a_*=0.5$
(dashed) and $a_*=0.998$ (solid). While this very simple model predicts 
frequencies which are fairly close to the observations, it is clear
that the gradient of this model in log space is 
different from that observed. 

Plainly the assumptions above are very simplistic. Global analytic
models of the hot flow with a standard $\alpha$ viscosity 
do not have $f_{visc}\propto f_K$ as they
depart from the self-similar solutions at $r<100$ due to the
requirement that the flow becomes supersonic (Narayan, Kato \& Honma 1997;
Gammie \& Popham 1998). Full numerical simulations also show that
$\alpha$ is not constant (e.g. Fragile et al 2007; 2009). Ingram \&
Done (2010) also suggest that $\zeta$ can change in neutron stars as
the material piles up onto a boundary layer. However, the similarity
between the mass scaled frequencies seen in neutron stars and black
holes shown in Fig.  \ref{fig:wkplot} make this now seem unlikely 
to be an important effect as it would not affect the black holes. 

Here then we simply assume that $\alpha (h/r)^2$ is a power law
function of radius, so that $f_{visc}=Br^{-m} f_K$. We choose values
for $B$ and $m$ which allow us to match the data in Figure \ref{fig:wkplot}.
We see good agreement with the observations for $B=0.03$ and $m=0.5$ (green
lines), again for $a_*=0.2$ (dot-dashed), $a_*=0.5$ (dashed) and $a_*=0.998$
(solid). We use this specific prescription for the viscous frequency in the
following section.

\section{The full model}  
\label{sec:mdot}

We consider a model where 
local fluctuations in the mass accretion rate of the
flow propagate down towards the central object (e.g. L97; K01).
Our method mainly follows that of AU06, with a few small differences.

We split the flow into annuli, characterized by a radius $r_n$ and
width $dr_n$, with logarithmic spacing so $dr_n/r_n$ is a constant for
all annuli from $r_o$ to $r_i$. As $f_{visc}\propto r^{-(m+3/2)}$ in
our prescription, constant $dr_n/r_n$ also implies constant $df/f$
which AU06 show is the feature required to produce a linear sigma-flux
relation. 

MRI fluctuations throughout the flow generate variability in all
quantities. Fluctuations in mass accretion rate will most likely
be damped by the response of the flow on timescales shorter than
the local viscous timescale. We therefore assume that the generated
power spectrum of mass accretion rate fluctuations at radius $r_n$
is given by a zero centred Lorentzian cutting off at the viscous
frequency
\begin{equation}
|\tilde{\dot{m}}(r_n,f)|^2 \propto {1\over 1+ 
(f/f_{visc}(r_n))^2 }
\label{eqn:poff}
\end{equation}
where $f_{visc}=-\frac{1}{R_g}v_r/r = Br^{-m}f_{K}$ as discussed at
the end of section \ref{sec:mod} and a tilde denotes a Fourier
transform. 

We start at the outermost annulus, so $r_1=r_o$, and generate the time
dependent fluctuations in mass accretion rate, $\dot{m}(r_1,t)$, from equation
\ref{eqn:poff} using the method of Timmer \& Koenig (1995). We normalize
each $\dot{m}(r_n,t)$ to have a mean of unity and fractional variability
$\sigma/I = F_{var}\sqrt{N_{dec}}$ where $F_{var}$ is the fractional
variability per decade in radial extent and $N_{dec}$ is the number of annuli
per decade in radial extent. Thus the mass accretion rate across the first annulus is 
$\dot{M}(r_1,t)=\dot{M}_0 \dot{m}(r_1,t)$ where $\dot{M}_0$ is the mean mass
accretion rate. This then propagates inward to the second annulus, travelling
a distance $dr_1$, which takes a time $t_{lag}=R_g~dr_1/v_r(r_1)$. When it arrives
at $r_2$, it has been filtered by the response of the flow which we take from
Psaltis \& Norman (2000) to get
\begin{equation}
\tilde{\dot{M}}_f(r_n,f) \propto \frac{\tilde{\dot{M}}(r_n,f)}{\sqrt{1+ [(dr_n/r_n)
(f/f_{visc}(r_n))]^2 }}.
\label{eqn:pandn}
\end{equation}
The mass accretion rate at the $n^{th}$ annulus is then given by
\begin{equation}
\dot{M}(r_n,t)=\dot{M}_f(r_{n-1},t-t_{lag})\dot{m}(r_n,t)
\end{equation}
where $t_{lag}=R_g~dr_n/v_r(r_n)$.
However, equation \ref{eqn:pandn} only filters out fluctuations on much shorter timescales
(by a factor $dr/r$) than the typical timescales generated in the annulus
(equation \ref{eqn:poff}) and so we can say
$\tilde{\dot{M}}_f(r_n,f)\approx \tilde{\dot{M}}(r_n,f)$ to a very good
approximation. The mass accretion rate at the $n^{th}$ annulus is therefore given
by
\begin{equation}
\dot{M}(r_{n},t)=\dot{M}(r_{n-1},t-t_{lag}) \dot{m}(r_{n},t),
\end{equation}
until the $N^{\rm th}$ annulus which is $r_i$. 

To transform this into a light curve requires an emissivity, $\epsilon
(r)$ such that the luminosity from each annulus is given by 
$dL(r,t)=1/2~\dot{M}(r_n,t) \epsilon(r) r_n dr_n c^2$
where we assume the emissivity $\epsilon(r) \propto
r^{-\gamma}b(r)$ where $b(r)$ is a boundary condition. 
For a Newtonian thin disc, $\gamma=3$ and we have the stress free
inner boundary $b(r)=3(1-\sqrt{r_n/r_i})$ but we note that the large
scale magnetic fields present in the large scale height flow can give
a stressed inner boundary condition $b(r)=1$ (Agol \& Krolik 2000; 
Beckwith Hawley \& Krolik 2008). We also allow
$\gamma$ to be a free parameter as the emission
need not exactly follow the radial dependence of gravitational energy
release as long as the \textit{total} energy release is gravitational.
A different emissivity for different energy bands gives a way
for the model to predict frequency dependent time lags
between hard and soft X-ray bands (K01; 
AU06). This will be the subject of a subsequent paper. 

\section{The fiducial model}
\label{sec:fiducial}

\begin{figure}
\centering$
\begin{array}{c}
\leavevmode  \epsfxsize=6.5cm \epsfbox{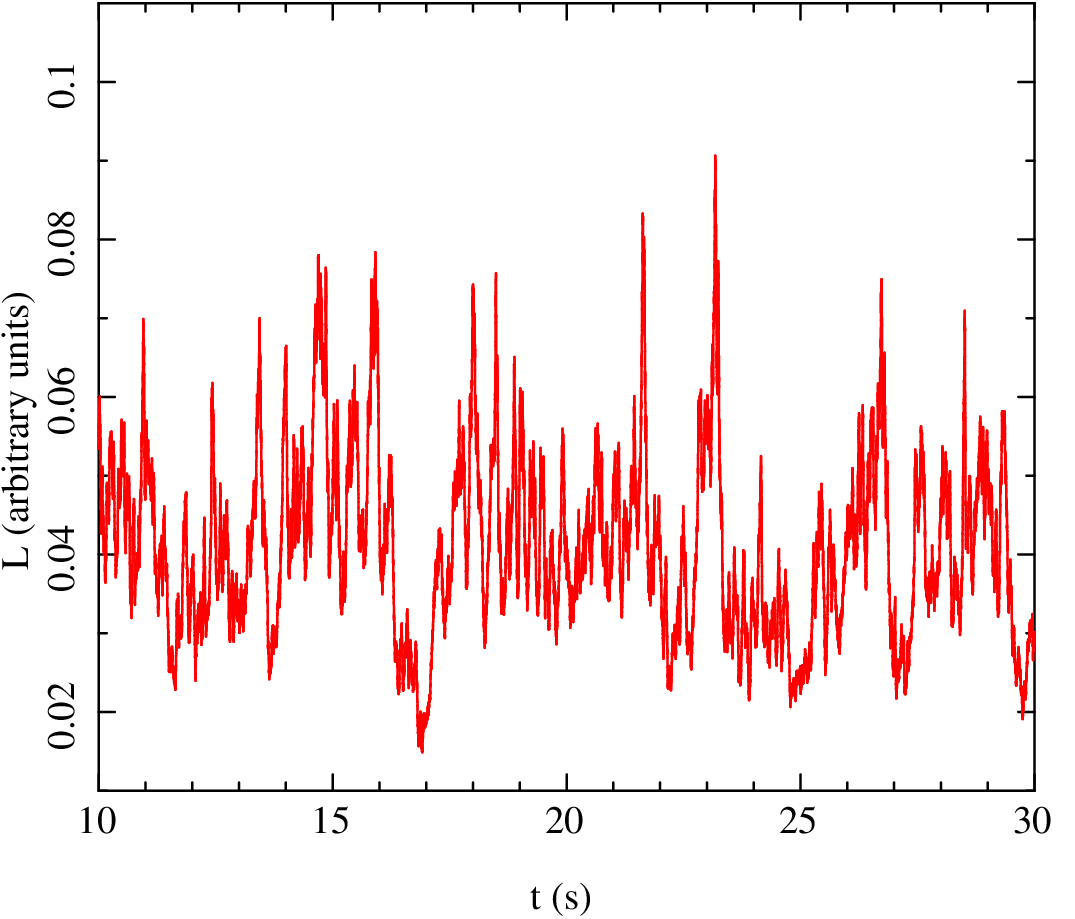}\\
\leavevmode  \epsfxsize=6.5cm \epsfbox{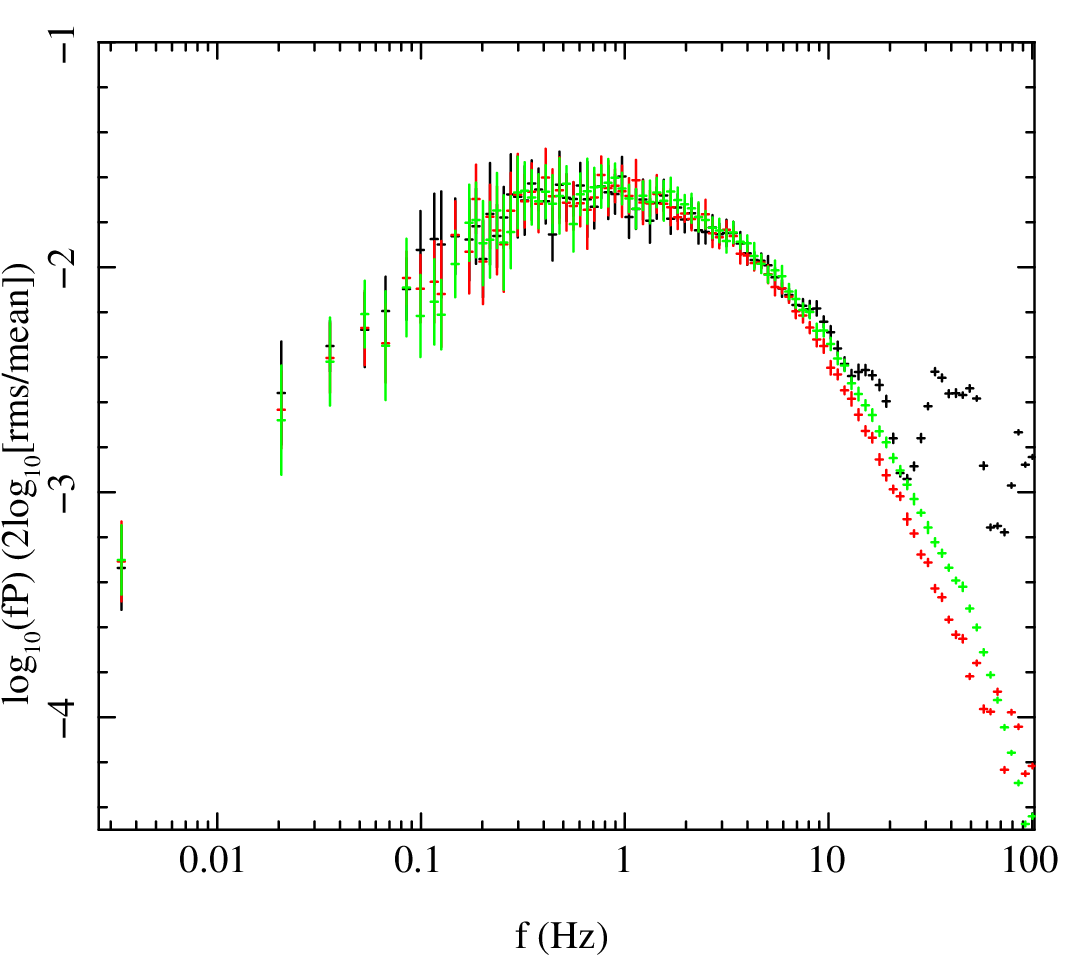}\\
\leavevmode  \epsfxsize=6.5cm \epsfbox{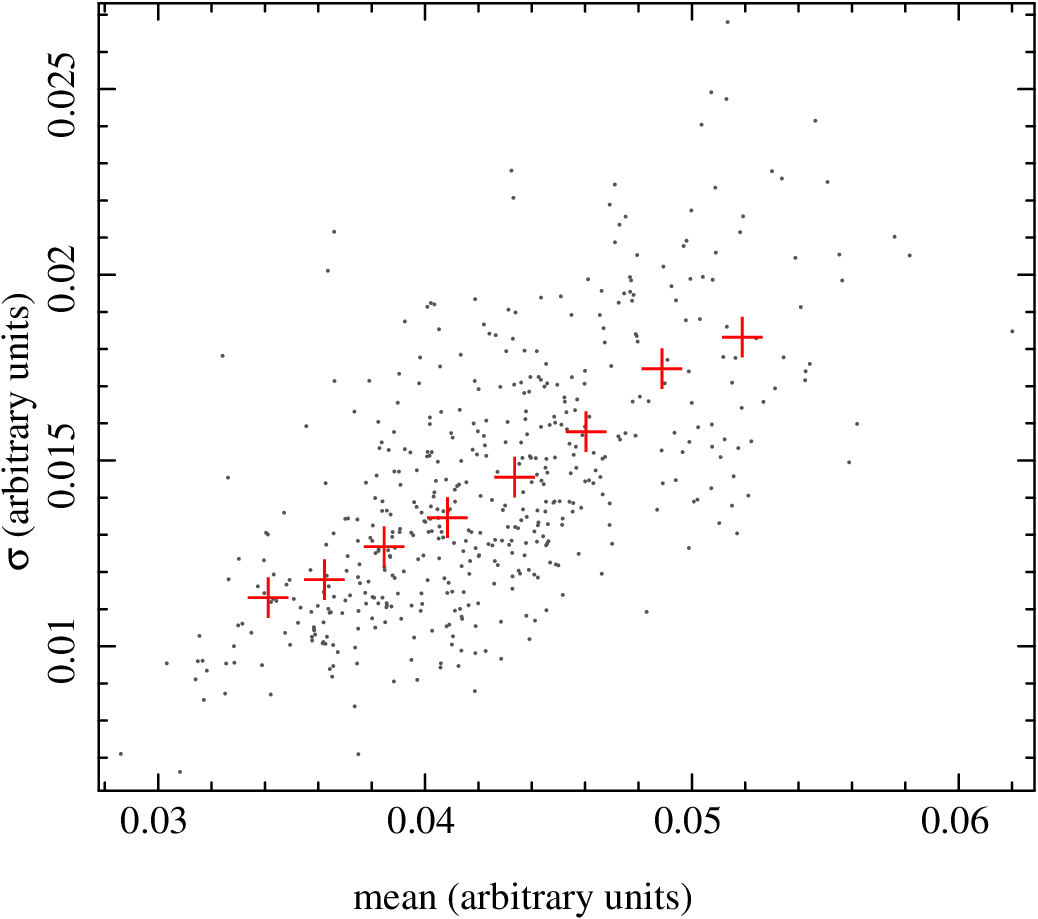}
\end{array}$
\caption{
\textit{Top (a):}
A 20 second section of the simulated light curve.
\textit{Middle (b):} The simulated power spectral density calculated
using 10 (black), 30 (red) and 100 (green) radial bins. We see
that, 10 bins is not enough to resolve the high frequency power
but 30 bins is a good approximation. 
\textit{Bottom (c):} The rms-flux relation for the light curve shown. We
see that this is linear as observed.
}
\label{fig:broadband}
\end{figure}

Figure \ref{fig:broadband}a shows 20~s of the resulting light curve
for a fiducial set of input parameters for a black hole mass of
$M=10M_{\sun}$ and a spin of $a_*=0.5$. We assume $r_{i}=2.5$, $r_o=20$,
$F_{var}=0.4$, $B=0.03$, $m=0.5$, $\gamma=4.5$ with a stressed inner
boundary condition (see Section \ref{sec:mod}).  We calculate the light curve with
$2^{22}$ time points, corresponding to $\sim 4096$~s (a typical length
for an RXTE observation) of data on a time binning of $0.00097$~s, and
30 radial bins.

Figure \ref{fig:broadband}b (red) shows the PSD of this light curve,
while the black and green points show the effect of changing the number
of radial bins to 10 and 100, respectively. Clearly, the high
frequency power is not well resolved with only 10 radial bins, while
the difference between 30 and 100 is very small. Hence in all that
follows, we use 30 radial bins for each simulation.

The PSD shows the same characteristic broadband noise features as are
seen in the power spectra of black hole binaries, namely band limited
noise, with low and high frequency breaks, peaking between 0.1-10~Hz.
Figure \ref{fig:broadband}c shows the rms-flux relation for the
fiducial light curve, derived from splitting this into 4s
segments. As with the data (Uttley \& McHardy 2001), we see a large
scatter before binning (gray points) but, after binning (red crosses),
we retrieve a linear flux-rms relation (AU06).  

\section{The truncated disc/hot inner flow model}
\label{sec:ro}

The major prediction of the truncated disc/hot inner flow model is
that the spectral softening as the source brightens from a low/hard
through to intermediate states is caused by the truncation radius of
the thin disc moving inwards (e.g. DGK07; Gierlinski, Done \& Page
2008). This radius also sets the outer edge of the hot flow, so this
predicts that $r_o$ decreases also. 

\begin{figure}
\centering
\leavevmode  \epsfxsize=6.5cm \epsfbox{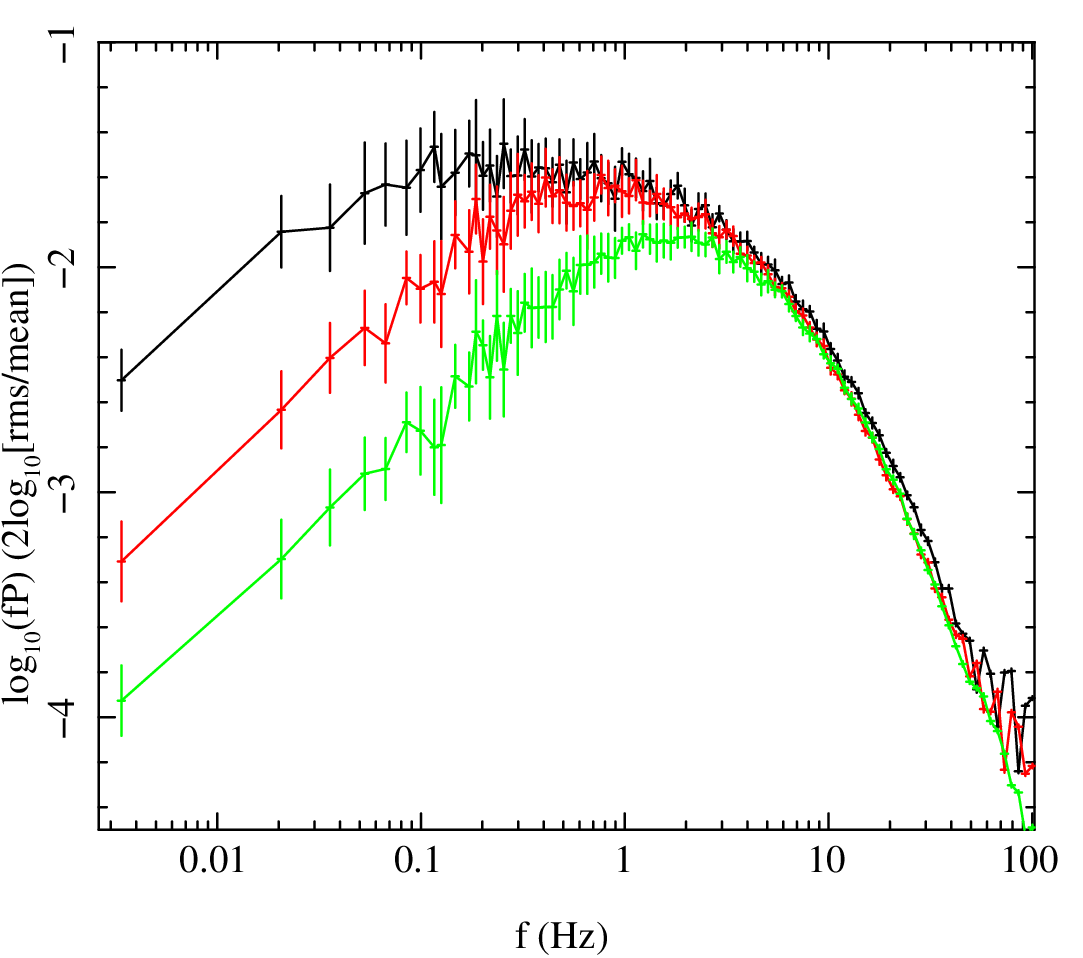}
\caption{PSD calculated using the fiducial parameters with
$r_o=50$ (black), $20$ (red) and $10$ (green) with total fractional
variability generated per decade in radius,  $F_{var}$, held constant.
This has the same characteristics as the
observed PSD of the data as the source softens from a low/hard to
intermediate state, namely that the low frequency power drops while
the high frequency power remains constant.}
\label{fig:fvar}
\end{figure}

Figure \ref{fig:fvar} shows the predicted PSD for
$r_o=50$, $20$ and $10$, as required to match the energy spectral
evolution (and low frequency QPO: IDF09), with all other parameters
held constant at the fiducial model values described above. 
The model {\em predicts} that decreasing the outer radius of the hot
flow leads to less low frequency power, while the high frequency power
remains constant. This is precisely what is seen in the PSD of the
data (DGK07; Gierlinski, Nikolajuk \& Czerny 2008). 

This is the first physical model of the power spectral behavior which
naturally reproduces the observations. The low frequency break is
close to the frequency of the viscous timescale at $r_o$, as proposed
by e.g. by Churazov et al (2001); Gilfanov \& Areief (2005); DGK07; Ingram
\& Done (2010).  However, the high frequency break is {\em not} at the
viscous frequency at $r_i$. We explore the origin of the high
frequency break below.

\subsection{Effect of propagation on the PSD shape}
\label{sec:cor}

\begin{figure}
\centering$
\begin{array}{c}
\leavevmode  \epsfxsize=6.5cm \epsfbox{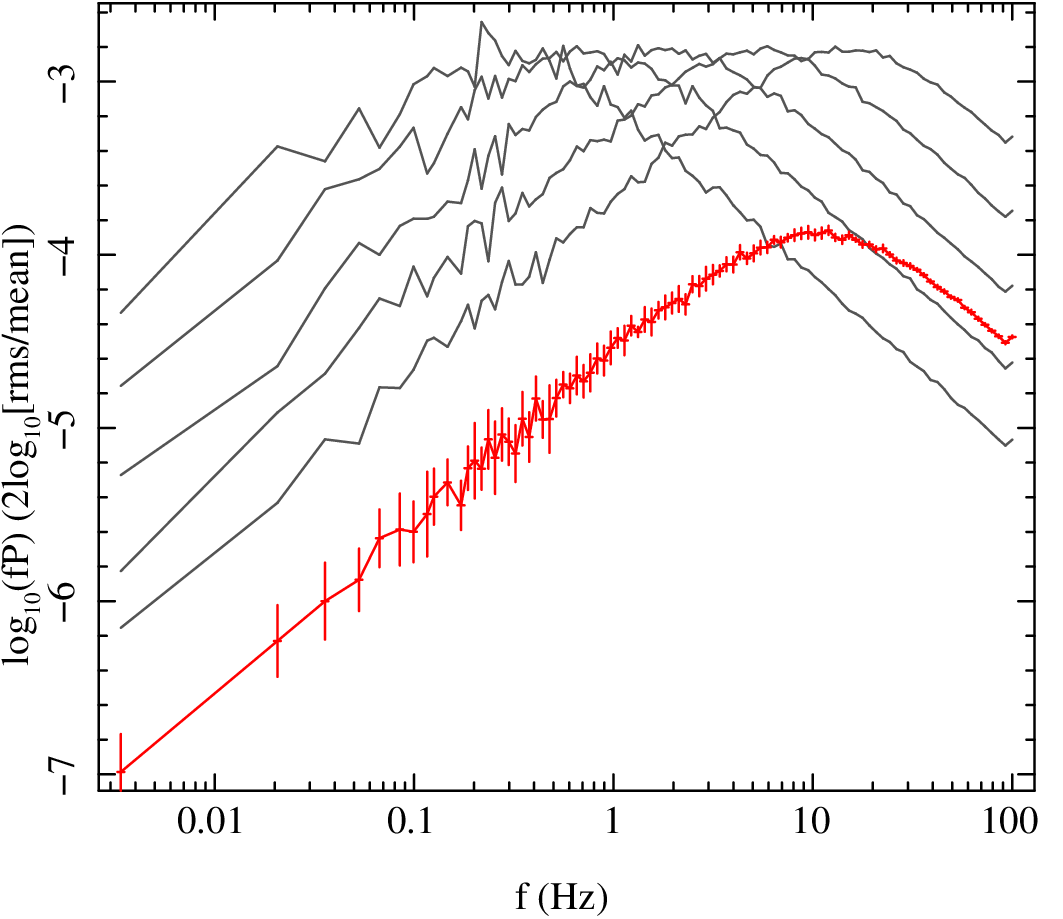}\\
\leavevmode  \epsfxsize=6.5cm \epsfbox{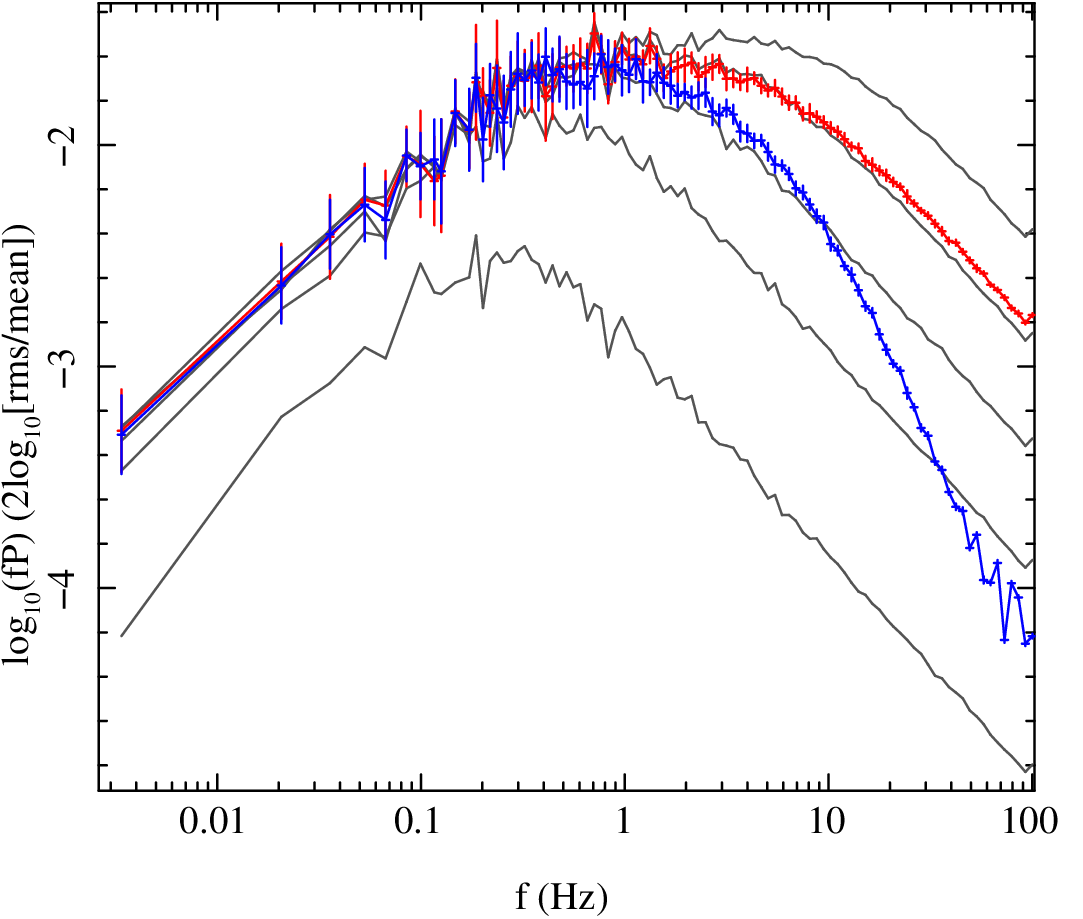}
\end{array}$
\caption{\textit{Top (a):} The gray lines are the power spectra of 5
simulated $\dot{m}(r_n,t)$ functions. We simulate 30 of these functions
but, for clarity, only plot 5 without showing the errors. The red line
is the PSD of the light curve created by assuming there to be no
propagation (i.e.$\dot{M}(r_n,t)=\dot{m}(r_n,t)$) and an emissivity index
of $4.5$. Because the functions we sum over are uncorrelated, the PSD of
the light curve looks like the (weighted) sum of the 100 PSDs with the
only difference being the normalization.  \textit{Bottom (b):} The gray
lines are now the power spectra of $\dot{M}(r_n,t)$ functions, i.e. we
now allow propagation. These are correlated at low frequencies but not
at high frequencies allowing the model to reproduce the observed
linear sigma-flux relation. The red line is the PSD of the light curve
if we do not consider the propagation time between annuli
(i.e. $t_{lag}=0$) and the blue line results if we do consider the
propagation time.  We see that the red line differs from the top plot
in 2 ways: the normalization is much higher and high frequency noise
is lost. However, much more high frequency noise is lost for the blue
line indicating that considering lags reduces high frequency
noise. These plots illustrate that the prediction from shot noise
models such as the top plot that the observed high frequency break is
the viscous frequency at the inner radius breaks down once we consider
a more advanced model capable of reproducing other observational
properties.}
\label{fig:cor}
\end{figure}

Figure \ref{fig:cor}a shows a model where the variability at each
radius is a Lorentzian at the local viscous frequency (see equation
\ref{eqn:poff}), but with no propagation so there is no causal
connection between annuli. We show the PSD of the resulting
$\dot{m}(r_n,t)$ functions from 5 of the 30 individual radial annuli,
from $r_o$ to $r_i$ as the gray lines on Figure \ref{fig:cor}a. These
peak, as expected, at $f_{visc}(r_o)$ and $f_{visc}(r_i)$.  The total
variability (red) is an emissivity weighted sum of these fluctuations,
but since they are uncorrelated, the effect of this is to strongly
dilute the total variability seen. This total PSD does have $f_h
\approx f_{visc}(r_i) \sim 12$Hz as our emissivity weighting strongly favors
the smallest radii, but $f_b > f_{visc}(r_o)$ ($\sim10$Hz and $\sim0.3$Hz
respectively). In fact, to achieve $fP_f \propto f^0$ as observed, we would
have to assume a completely flat emissivity profile, which seems very
unlikely. More fundamentally, such uncorrelated fluctuations {\em cannot}
reproduce a linear rms-flux relation.

This is in sharp contrast to a model where fluctuations propagate down
in radius. The resulting PSD from the same set of radii are shown in
Figure \ref{fig:cor}b, where the power in each annulus increases
strongly with radius as the MRI power generated in each annulus is
modulated by the propagating fluctuations from all radii prior to it.
The red line shows the resulting emissivity weighted power spectrum
from the total flow assuming that time lags between radii are
negligible. This preserves the maximum correlation between variability
at different annuli i.e. gives the least dilution between fluctuations
in different annuli. This is very different to that in Figure
\ref{fig:cor}a, both in normalization and shape. The normalization is
dramatically enhanced because the long timescale fluctuations are
correlated together, so at low frequencies
the power from different radii add together as
they are in phase. This gives $f_b\approx f_{visc}(r_o) \sim 0.3$Hz
as the correlated variability weighting to larger radii is stronger
than the emissivity weighting to smaller radii. 
However, at the fastest timescales, the power is
mainly generated at the smallest radii, so it does not correlate with
any other fluctuations generated at larger radii, so is not enhanced
in the same way.

The blue line shows how time delays dramatically change the high
frequency break as the propagation time prevents the mass accretion
rate from two consecutive annuli from being correlated on time scales
shorter than $t_{lag}$. This reduces the correlation
between the fastest timescale variability, strongly suppressing high
frequency power. Thus in the propagating fluctuation model, the low
frequency break is $f_b \approx f_{visc}(r_o)$ but $f_h <<
f_{visc}(r_i)$.

\subsection{Emissivity and boundary condition}
\label{sec:boundary}

\begin{figure}
\centering
\leavevmode  \epsfxsize=6.5cm \epsfbox{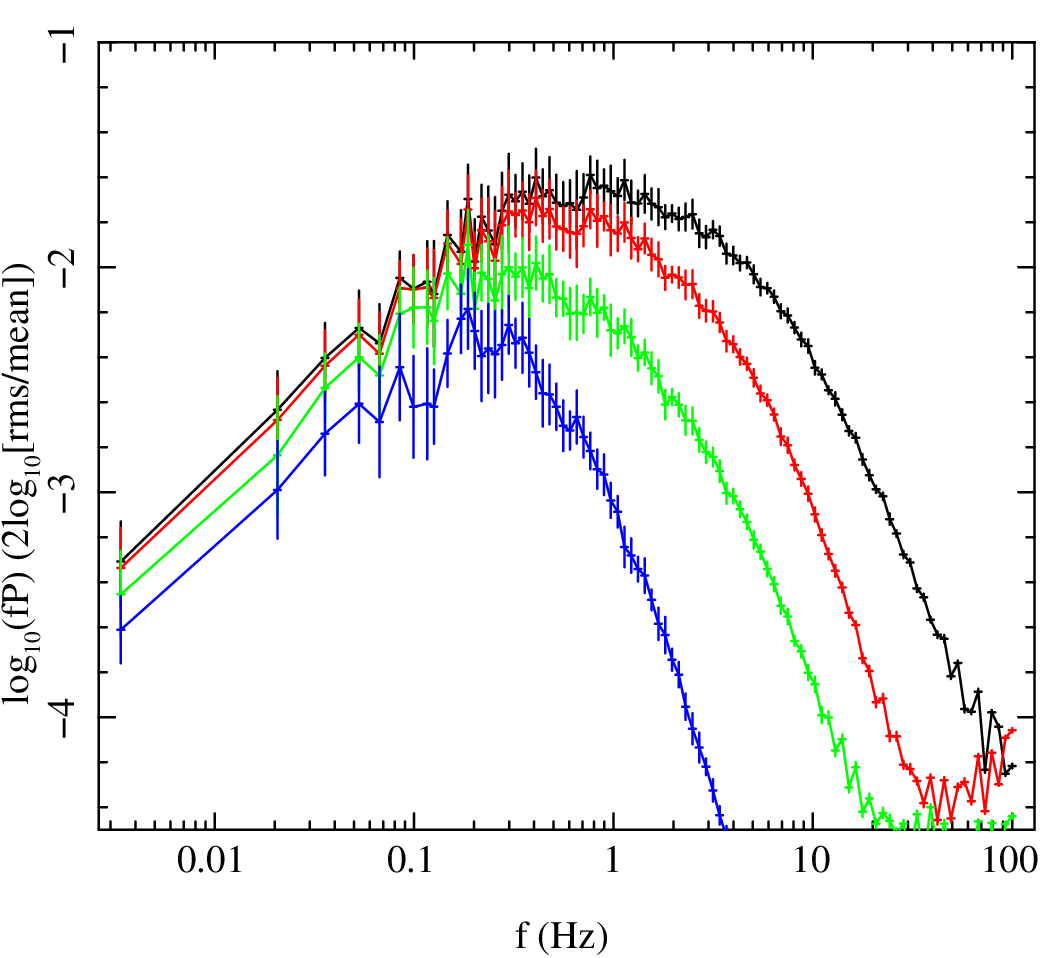}
\caption{
The PSD calculated assuming $b(r)=1$ and $\gamma=4.5$ (black),
$b(r)=$stress free and $\gamma=4.5$ (red), $b(r)=$stress free
and $\gamma=3$ (green), all with $r_i=2.5$. The blue points are
for $b(r)=$stress free, $\gamma=3$ and $r_i=6$. This illustrates
that we can reduce the predicted high frequency noise by changing
boundary condition, emissivity index or inner radius.
}
\label{fig:boundary}
\end{figure}

We use an emissivity to translate the fluctuations in mass accretion
rate to a luminosity. This emissivity is in two parts, firstly a power
law dependence in radius, and secondly a boundary condition.  Our
fiducial model parameters have $\gamma=4.5$ and a stressed boundary
condition, $b(r)=1$. This emissivity peaks at $r_i$, so 
fluctuations from the very smallest radii are
given most weight.

Figure \ref{fig:boundary} compares this (black line) with results
using the same power law radial dependence, but with 
a stress--free inner boundary condition (red line),
$b(r)=3(1-\sqrt{r/r_i})$. This emissivity goes to zero at the
innermost radius, so the highest frequency fluctuations are
strongly suppressed. However, this also has a more subtle effect
on the region between the two breaks, as there is a
gradual decrease in weighting of fluctuations 
below $r=2r_i$, and a stronger weighting to the fluctuations at larger
radii, giving the tilt between $f_b$ and $f_h$. 

This effect is similar to that of changing the radial dependence of
the emissivity. The green line shows $\gamma=3$ with a stress-free
boundary condition, showing an even stronger tilt to the PSD between
$f_b$ and $f_h$ (green). However, it is also similar to changing the inner
radius of the flow. The blue line in Figure \ref{fig:boundary}
shows the resulting PSD from $\gamma=3$ and a stress-free inner
boundary condition with $r_i=6$. 
Thus there are degeneracies between the two parts to the emissivity
and the inner radius, making it unlikely that they can all be
uniquely constrained by the observed PSD.

It is clear from this analysis that 
while the low frequency break is fairly strongly linked to the viscous
timescale of the outer radius of the hot flow (as assumed in section
\ref{sec:mod}), the high frequency
break is rather more complex, depending on propagation correlations,
emissivity, boundary condition and inner radius in addition to the
viscous timescale. This makes it
difficult to directly associate the high frequency break with any
physical parameter of the models. Instead, we now use the additional
information from the QPO to remove some of the degeneracies inherent
in this model for the broadband noise.

\section{The QPO: Precession and surface density}  \label{sec:sigma}
\label{sec:qpo}

For our fiducial model, we used the observed relation between the low
frequency break and LF QPO to set the radial dependence of the viscous
timescale, assuming that the low frequency break was set by the viscous
timescale at $r_o$ and that the QPO was Lense-Thirring precession of the
entire hot flow from $r_o$ to $r_i$ (Section \ref{sec:mod}). This assumed
that the surface density of the hot flow, $\Sigma=\Sigma_0 r^{-\zeta}$
between $r_o$ and $r_i$, with $\zeta=0$. However, the broadband noise
model described above {\em calculates} a self-consistent 
surface density as mass conservation implies
\begin{equation}
\dot{M}(r_n,t)=-2\pi r_{n} v_r(r_{n}) \Sigma(r_{n},t),
\end{equation}
(Frank, King \& Raine 1992) where $v_r$ is now expressed in units of $c$,
$\dot{M}$ in units of $\dot{M}_0$, $\Sigma$ in units of $\dot{M}_0/(cR_g)$
and $r$ in units of $R_g$. Using our velocity prescription, we can then
easily show
\begin{equation}
\Sigma(r_n,t) = \frac{\dot{M}(r_n,t) r^{m-1/2}}{B} .
\label{eqn:sigzet}
\end{equation}
This means that, for the time averaged surface density, $\zeta = m-1/2$
giving extra physical motivation for the parameters
used in section \ref{sec:mod} ($\zeta=0$ and $m=1/2$).

\begin{figure}
\centering$
\begin{array}{c}
\leavevmode  \epsfxsize=6.5cm \epsfbox{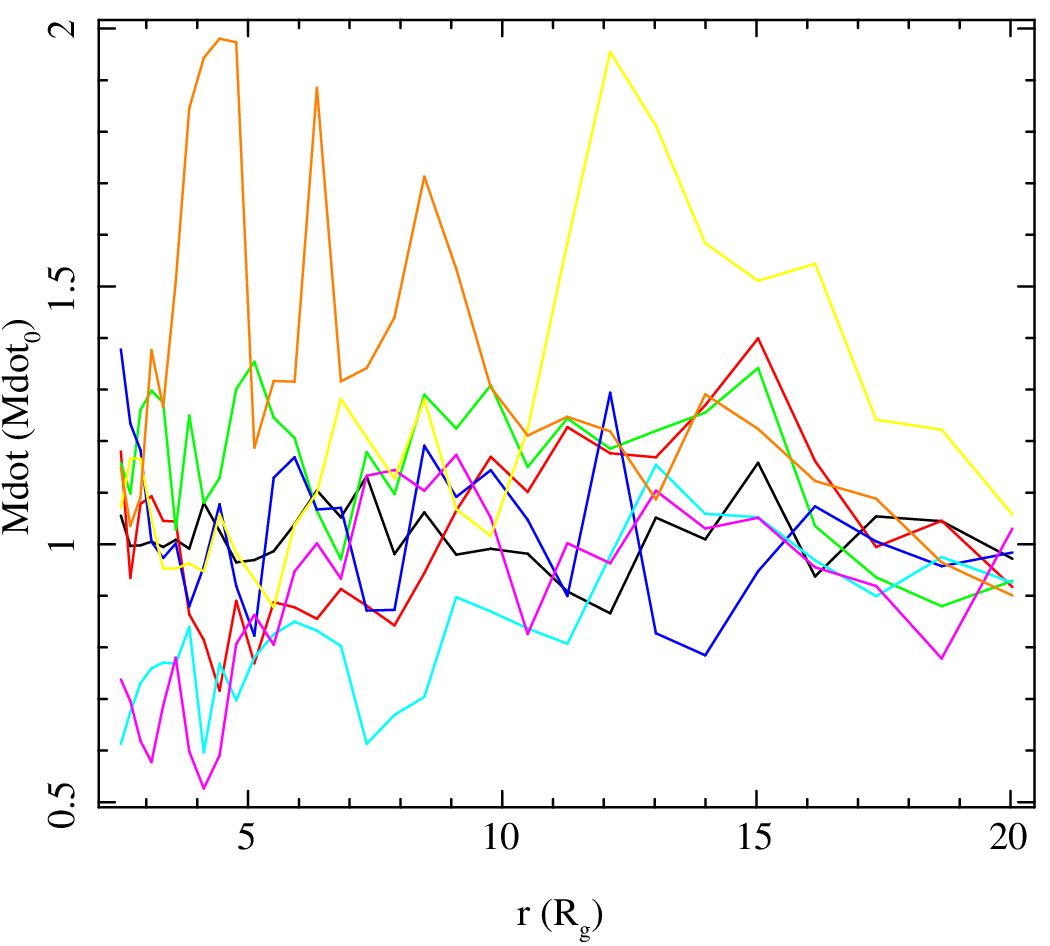}\\
\leavevmode  \epsfxsize=6.5cm \epsfbox{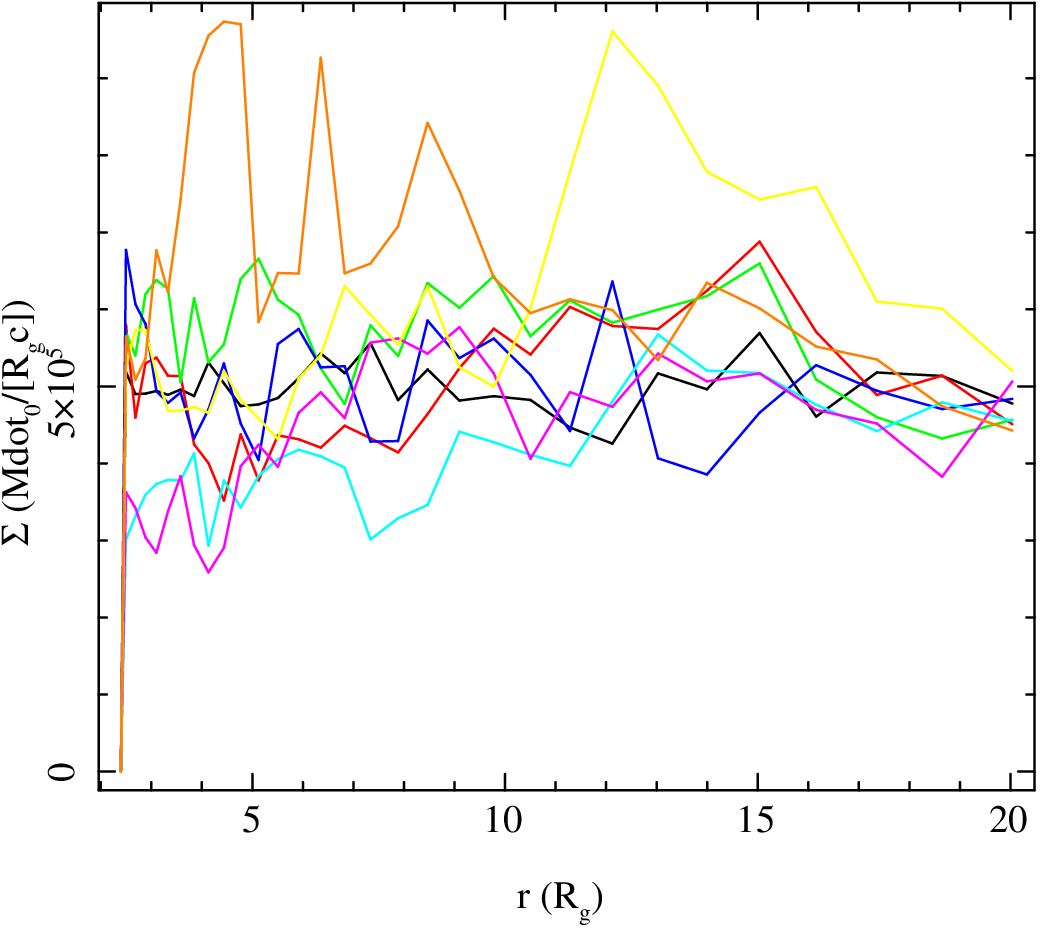}
\end{array}$
\caption{\textit{Top:} Mass accretion rate as a function of radius shown here
at a number of different times.
\textit{Bottom:} Surface density as a function of radius shown at the same
times as the lines of corresponding colour in the top plot. This is calculated
by applying mass conservation in the flow. The dotted line in the bottom
plot represents the bending wave radius for $a_*=0.5$ and $h/r=0.2$ which
seems to trace the point where the surface density begins to drop off.}
\label{fig:surfaced}
\end{figure}

Figure \ref{fig:surfaced} shows $\Sigma(r_n)$ plotted at a number of
different times ($0,256,512,...,1792~s$) along with the corresponding
$\dot{M}(r,t)$ function. $\dot{M}(r,t)$ is quite clearly more variable
at small $r$. This is because we have assumed the variability
\textit{generated} in each logarithmic annulus to be the same but
annuli at smaller radii include also the fluctuations that have
propagated down from large $r$ and so the \textit{emitted}
variability is greater (see Figure \ref{fig:cor}b). We do not see a
drop off in surface density at the bending wave radius like that
seen in simulations (e.g. Fragile 2009) because we assume that the infall
velocity can be given by a power law. It is clear that, for the surface
density drop off at a given point, the infall velocity must accelerate at
that point. In a future paper, we will investigate this model with a more
advanced velocity prescription.

Therefore the broadband noise model above, set by $r_i$, $r_o$ and
$F_{var}$ {\em predicts} the QPO frequency at any point in time.  The
fluctuations in surface density with time {\em predict} that the QPO
frequency changes, i.e. it is quasi--periodic rather than truly
periodic. However, the precession frequency will not respond
instantaneously to these changes, as their effect is only communicated
across the entire hot flow by bending waves. These travel at the sound
speed, faster by a factor $\sim \alpha$ than the viscous timescale
across the region, so we calculate the QPO frequency every $\sim 4$~s
rather than at every point. We then average these values to get the
predicted QPO frequency, $f_{QPO}$, and use the dispersion around this
to set the {\it r.m.s.} variance of these QPO frequencies,
$\sigma_{QPO}$. 

\begin{figure}
\centering
\leavevmode  \epsfxsize=6.5cm \epsfbox{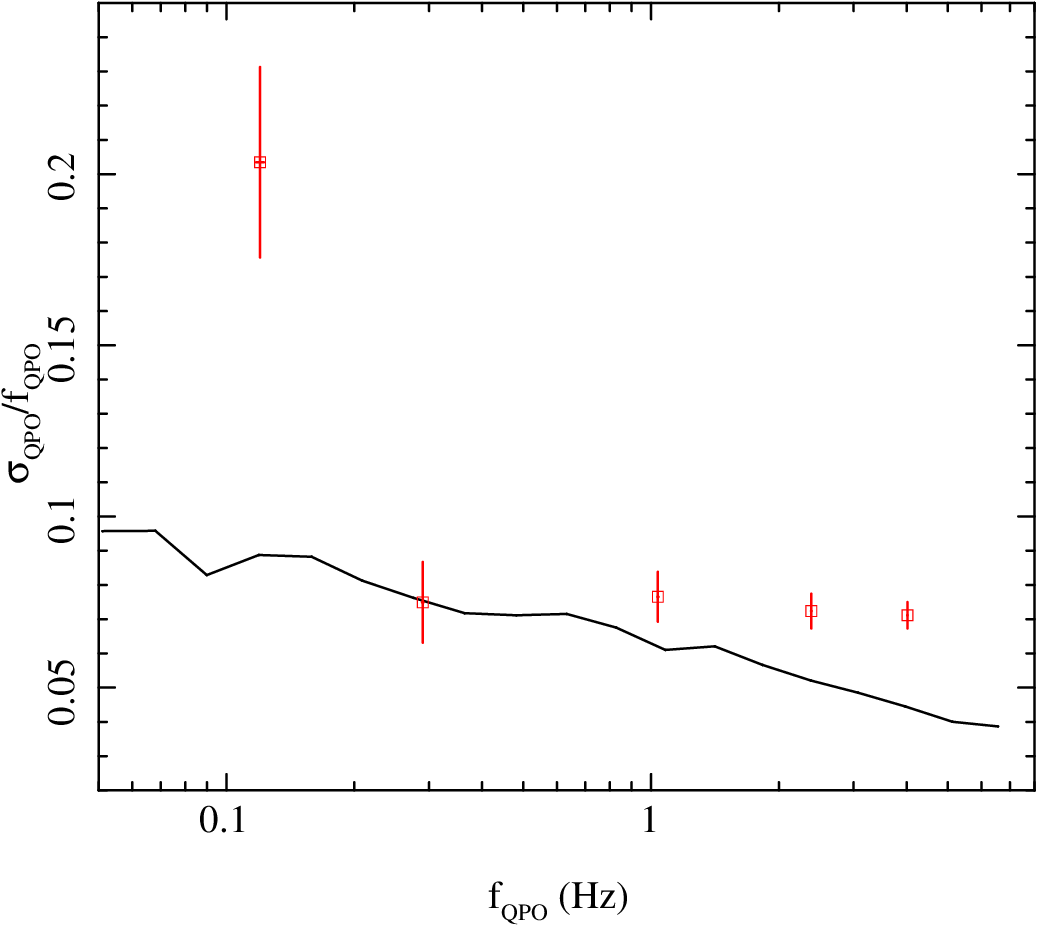}
\caption{
Fractional variability of the precession frequency plotted against the
average precession frequency (black line). These are calculated by measuring
the precession frequency every $4$s for a number of different truncation radii,
ranging from $300-10$, and taking the average and standard deviation over a
2048s duration. The red squares show the observed QPO width and frequency in
data from the 1998 rise to outburst of XTE 1550-564. We see broad agreement with
the data, however, other effects such as on-time of the QPO can decrease the
coherence of the signal so we note that we are only able to predict a lower
limit for the width of the QPO}
\label{fig:qvsqpo}
\end{figure}

Fig. \ref{fig:qvsqpo} shows $\sigma_{QPO}/f_{QPO}$ as
a function of $f_{QPO}$ as $r_o$ varies from 300--10 in the fiducial
model.  This decrease in $r_o$ not only leads to an increase in QPO
frequency, but also to a decrease in the QPO width, or equivalently, an
increase in its coherence/quality factor $Q=f_{QPO}/\sigma_{QPO}$.
This correlation is well known in QPO data from BHB (e.g. Belloni, Psaltis
\& van der Klis 2002; Rao et al 2010). Our model
provides the first physical explanation for this effect as the smaller
the radial extent, the higher the QPO frequency, but also the smaller
the fluctuation power, giving smaller jitter in frequency.  The red
squares in Fig \ref{fig:qvsqpo} show the {\em observed} frequency and
width of the QPO from data from the 1998 rise to outburst of the BHB
XTE J1550-564 (see section 7). The model matches the trend
in the data fairly well, and forms a lower limit to the width of the
QPO. However, other effects such as the on-time of the QPO (see
Lachowicz \& Done 2010) can decrease the coherence of the signal, so
here we simply use the data to determine the QPO width (parametrized
as a quality factor $Q=f_{QPO}/\sigma_{QPO}$) after using the
broadband noise surface density to determine the QPO frequency,
$f_{QPO}$.

\begin{figure}
\centering$
\begin{array}{c}
\leavevmode  \epsfxsize=6.5cm \epsfbox{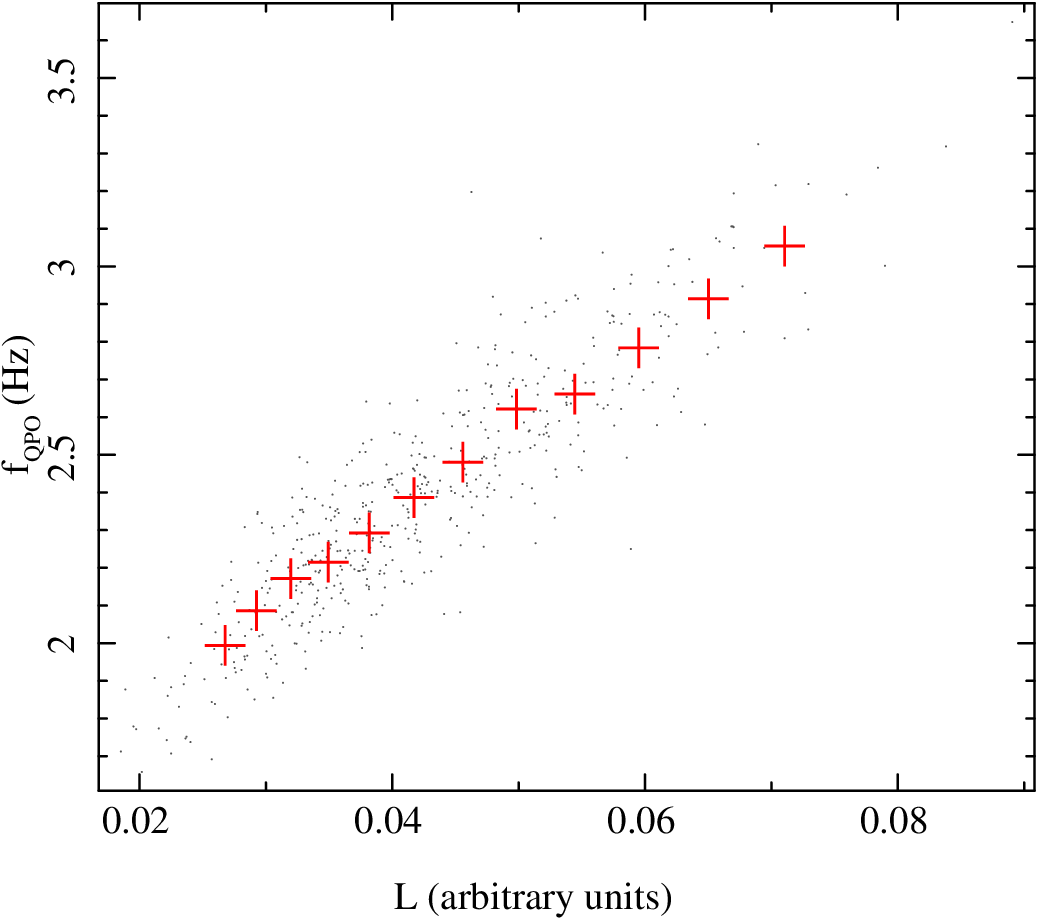}\\
\leavevmode  \epsfxsize=6.5cm \epsfbox{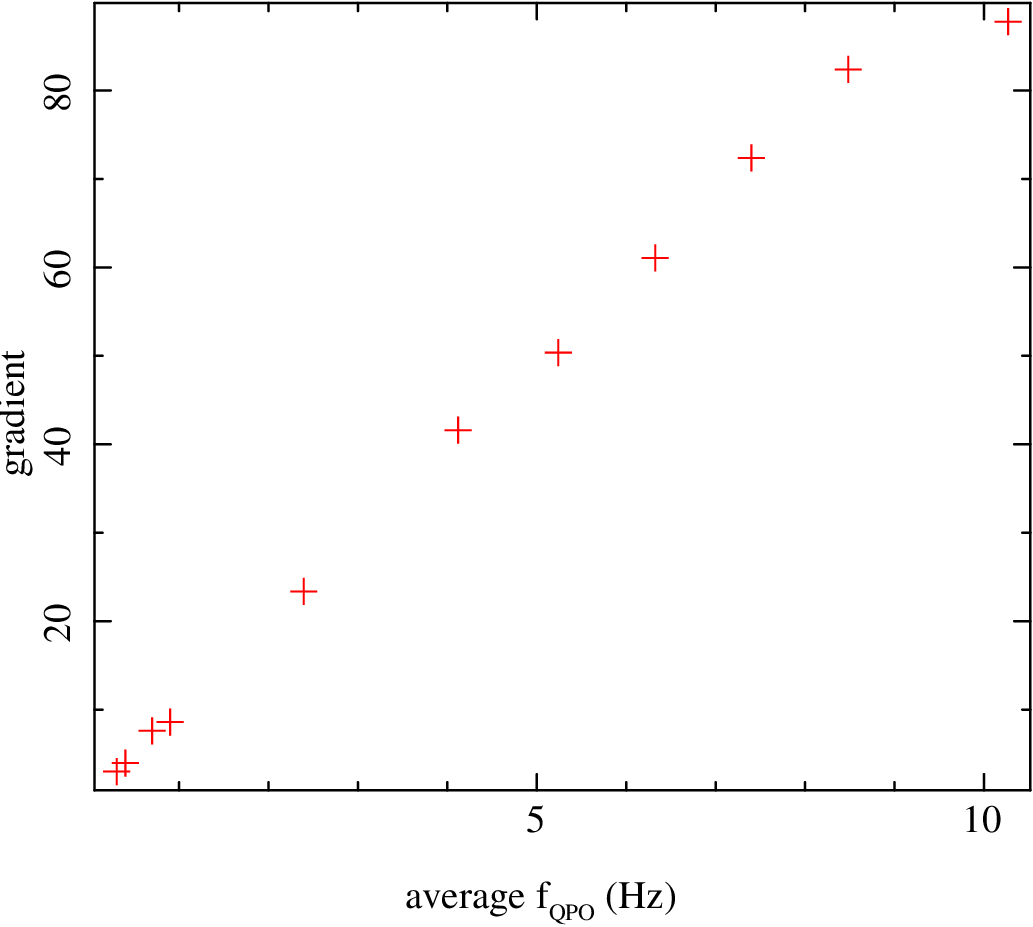}
\end{array}$
\caption{
Precession frequency plotted against luminosity, where both are calculated
at $4$s intervals using the fiducial model parameters (gray points). After
binning (red crosses), we clearly see a linear relationship between the
two quantities. This relationship has recently been discovered in data from
the 1998 rise to outburst of XTE J1550-564 (Heil et al in print), demonstrating
the substantial predictive power of this model.}
\label{fig:lvsqpo}
\end{figure}

\begin{figure}
\centering
\leavevmode  \epsfxsize=6.5cm \epsfbox{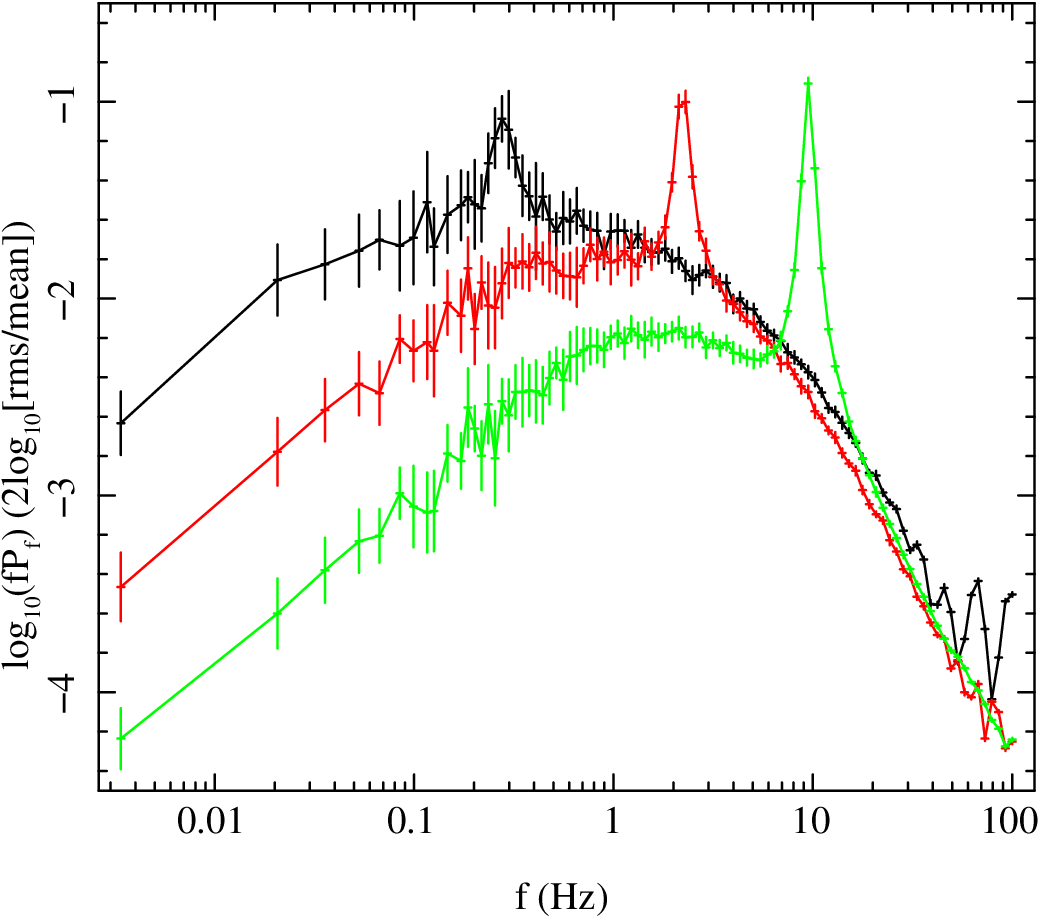}
\caption{
The full PSD calculated using the fiducial model parameters with $r_o=50$ (black),
$20$ (red) and $10$ (green). Here, the QPO is represented by a Lorentzian
centred at the precession frequency with the width set by the r.m.s variance
in precession frequency (see text).}
\label{fig:qpoandbb}
\end{figure}

The model also predicts another correlation, one between the QPO
frequency and flux on short timescales.  The top plot of Figure
\ref{fig:lvsqpo} shows this for the fiducial model (i.e. $r_o=20$),
with precession frequency calculated every $4$s together with the
instantaneous luminosity at that time.  After binning (red crosses),
there is a clear linear relation between the two. This happens because
both the QPO frequency and the luminosity depend on the mass accretion
rate fluctuations.  A perturbation in mass accretion rate at large
$r$ will lead to a perturbation in the surface density. This will
reduce the precession frequency but will have little effect on the
luminosity because the emissivity is quite steeply weighted towards
small $r$. Later on, this perturbation will have propagated inwards
to small $r$ where it has the effect of increasing the precession
frequency, but now also has much more of an effect on the luminosity.
Heil et al (2011) have recently discovered this correlation in data
from the 1998 rise to outburst of XTE J1550-564 (the same data as we
will be considering in section \ref{sec:data}). They also find that
the gradient of this relation is steeper for observations with a
higher QPO frequency. This is also predicted by the model as
illustrated in the bottom plot of Figure \ref{fig:lvsqpo} where we
have measured the gradient of the $f_{QPO}$-$L$ relation and the
average QPO frequency for $11$ different $r_o$ values. There is
clearly a very strong correlation as is seen in the data. This is
because an \textit{absolute} change in precession frequency depends
on a \textit{fractional} change in mass accretion rate whereas an
\textit{absolute} change in luminosity depends on an \textit{absolute}
change in mass accretion rate. The same absolute change in mass
accretion rate at a given radius and time constitutes a larger
\textit{fractional} change for small $r_o$ than for high $r_o$.
Therefore the luminosity will experience exactly the same change in
both instances but the precession frequency will undergo a larger
change when $r_o$ is smaller. The fact that these are
\textit{predicted} properties of the model constitutes strong support
for its validity.

In the following section, we will fit the PSD from our model to
data. For this, we will need to predict a \textit{shape} for the QPO
light curve as well as a frequency.  The data show that the QPO has a
power spectrum which can be represented by a Lorentzian at the
fundamental frequency, $f_{QPO}$, together with its second and third
harmonic and sub-harmonic i.e. at $2f_{QPO}$, $3f_{QPO}$ and
$1/2f_{QPO}$ (e.g. Belloni, Psaltis \& van der Klis 2002). Our model for
the QPO in terms of Lense-Thirring precession {\em predicts} the shape
of the modulation of the emission from the hot flow via variation of
projected area, self-occultation
and seed photons (IDF09). We will explore this further in a later
paper (Ingram, Done \& $\dot{Z}$ycki in prep), but here we simply
assume that all the harmonics have the same quality factor, $Q$,
and allow the power in each harmonic to be a free
parameter.  We then generate a QPO light curve, $L_{QPO}$, using these
narrow Lorenzians as input to the Timmer \& Koenig (1995) algorithm,
and add this to the light curve already created for the broadband
noise.

We show an example of the final predicted PSD in Figure
\ref{fig:qpoandbb}, using the fiducial model parameters with $r_o=50$
(black), $20$ (red) and $10$ (green).  For clarity, we have set the
normalization of all the harmonics other than the fundamental to $0$,
set the width of the QPO using the model prediction of
$\sigma_{QPO}/f_{QPO}$, and set its rms power $\propto 1/r_o$. These
PSD show all the main features seen in the data during spectral
transitions of BHB (Gierlinski, Nikolajuk \& Czerny 2008).

%==========================================
\section{Fitting to XTE J1550-564} \label{sec:data}
%==========================================

We now have a model which can produce both the broad band noise and
QPO self consistently from the same changing geometry as required by
the corresponding evolution of the energy spectra, namely a changing
outer radius of the hot flow (set by the changing inner radius of the
thin disc). This model gives the major correlation between the low
frequency break and low frequency QPO. The propagating fluctuations
naturally give the rms-flux relation, while the QPO model also
gives a framework in which to understand the increase in coherence of
the QPO with frequency, and even the short timescale correlation of
the QPO frequency with flux. Plainly, the next step is to use this
model to fit real data.

We use RXTE data from the 1998 outburst of XTE J1550-564 (Remillard et
al 2002; Sobczak et al 2000; Rao et al 2010; Wilson \& Done 2001). We
look at 5 specific observations with observational IDs: 30188-06-03-00,
30188-06-01-00, 30188-06-01-03, 30188-06-05-00 and 30188-06-11-00;
hereafter observations 1-5 respectively. We only consider energy
channels 36-71 (corresponding to 10-20~keV) in order to avoid disc
contamination. 

In all our results so far we have shown power spectra produced by
averaging the logarithm of the periodogram as Papadakis \& Lawrence
(1993) show that, for a red noise variability process, the error
estimate converges to Gaussian with only $\sim 20$ samples, rather than
the $\sim 50$ required for averaging the linear power estimates. Our
simulations average over $M=32$ samples of the PSD but the number of
segments from the real data cannot be chosen as they are simply set by
the length of the available data. These give $M=41,~26,~13,~14$ and $14$
respectively, implying that the errors on the PSD from the data are
only approximately Gaussian. A further complication is that the
error distribution of the periodogram is dependent on the shape of
the underlying power spectrum
(e.g. Mueller, Madejski \& Done 2004; Mueller \& Madejski 2009)
meaning that even $M=20$ may not be
enough to give Gaussian errors in this particular case. Nonetheless, we
choose to use $\chi^2$ minimization for simplicity, but we then do
{\em a posteriore} checks on the goodness of fit using the rejection
probability method of Uttley et al (2002) and Markewicz et al (2003).
This entails first calculating
\begin{equation}
\chi^2_{dis}=\frac{1}{M}\sum_{k=1}^{M} \sum_f
\frac{(\overline{P_{sim}(f)}-\log_{10}[P_{obs,k}(f)])^2}{(\overline{\Delta P_{sim}(f)})^2}
\end{equation}
where $\overline{P_{sim}(f)}$ is the logarithmically smoothed simulated
periodogram, with error $\overline{\Delta P_{sim}(f)}$, to be compared to
$M$ raw observed periodograms, $P_{obs,k}(f)$. We then re-simulate many
(512) light curves and calculate many different values of
\begin{equation}
\chi^2_{dis}(sim)=\frac{1}{M}\sum_{k=1}^M \sum_f
\frac{(\overline{P_{sim}(f)}-log_{10}[P_{sim,k}(f)])^2}{(\overline{\Delta P_{sim}(f)})^2}.
\end{equation}
The rejection probability, $P_{rej}$, is given by the percentile of $\chi^2_{dis}(sim)$ values
which are lower than the one $\chi^2_{dis}$ value. For example, if 128 of the
$\chi^2_{dis}(sim)$ values were smaller than the one $\chi^2_{dis}$ value,
the rejection probability is $25\%$. This means there is a $25\%$ chance that
the the observed light curve was generated by a different underlying variability
process than the model.

One aspect of using real data is that there are 
uncertainties associated with each point on the light curve. 
Since we use the logarithmically averaged PSD we need the power
to be positive definite so we cannot do 
white noise subtraction of the data PSD. Instead we 
add white noise to the simulation, and compare this
with the total (including noise) power spectrum of the real data
(see also Uttley et al 2002). 

We incorporate our model for the power spectrum into {\sc xspec},
using the local model functionality. Our model outputs the
logarithmically averaged total (white noise included) periodogram as a
function of Fourier frequency rather than the more familiar flux as a
function of energy. We then fit this to the similarly logarithmically
averaged total power spectra from the real data. The model parameters
are the physical parameters governing the behavior of the flow. The
most fundamental of these is the relation governing the viscous
frequency at any radius. We have parametrized this as a power law
function of the Keplarian frequency at each radius,
$f_{visc}=Br^{-m}f_{kep}(r)$, but the parameters $B$ and $m$ are
unknown, so are free in the fits, as is $F_{var}$, the intrinsic level
of the MRI fluctuations in mass accretion rate.  Then there are the
geometric parameters $r_o$ and $r_i$, where $r_i$ also couples to the
emissivity, both in terms of the index $\gamma$ and the boundary
condition (either stressed or stress free).  The combination of $r_o$,
$r_i$ and the surface density profile fixes the QPO frequency, but we
leave its width (parametrized as a quality factor, Q) and
normalization as free parameters. We also allow the normalization of
the second, third and subharmonic to be free parameters, but fix these
to have the same Q as that of the fundamental. The final free
parameter is the normalization of the simulated white noise level,
$\sigma_{lc}$, which we then subtract from both data and model in
order to plot results.

The truncated disc model predicts that $r_o$ is the major parameter
which changes across this transition, where $r_i$ remains fixed.
Hence we fit all 5 power spectra simultaneously in {\sc xspec}, with
$r_i$ tied across all 5 datasets. Because of the correlation between
$r_i$ and both parts of the emissivity (see section \ref{sec:boundary}),
we choose to compare results with some of these parameters fixed at
different values. 

\subsection{Fit results}
\label{sec:results}

\subsubsection{Stressed inner boundary}
\label{sec:stressed}

\begin{figure}
\centering$
\begin{array}{c}
\leavevmode  \epsfxsize=5.5cm \epsfbox{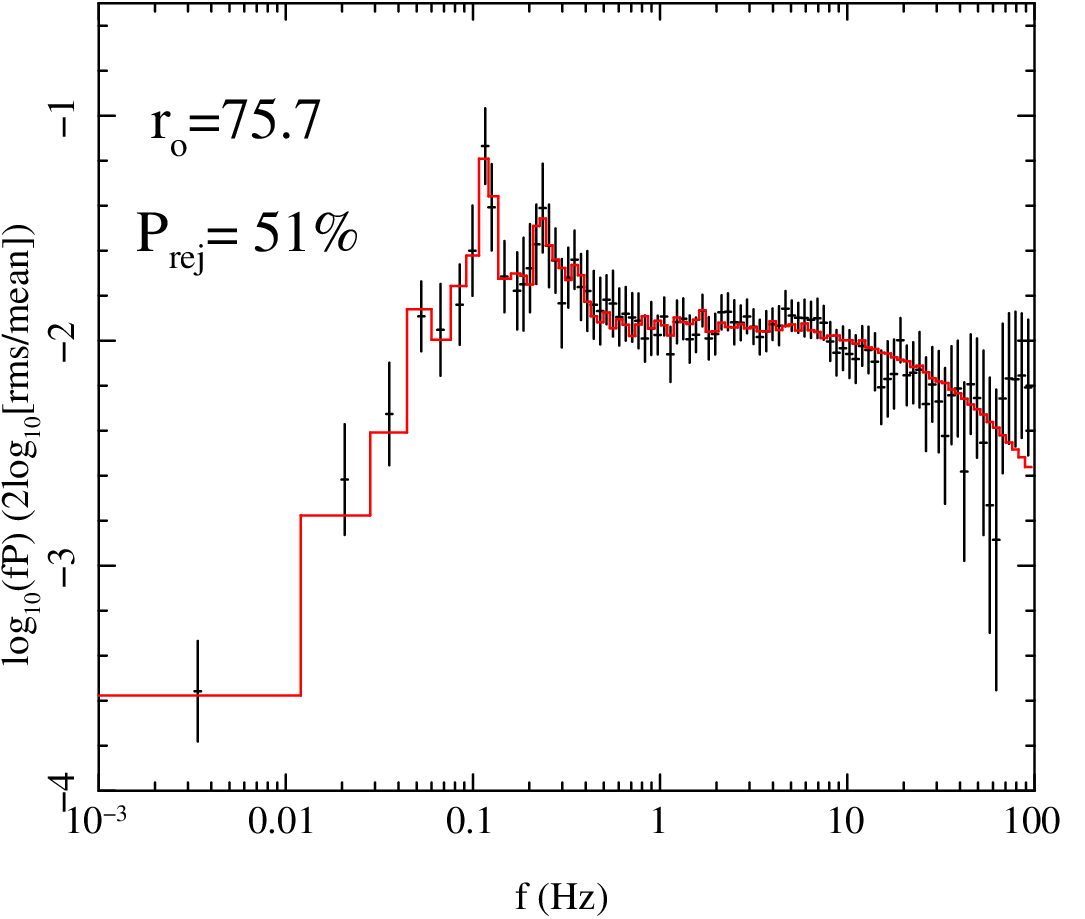}\\
\leavevmode  \epsfxsize=5.5cm \epsfbox{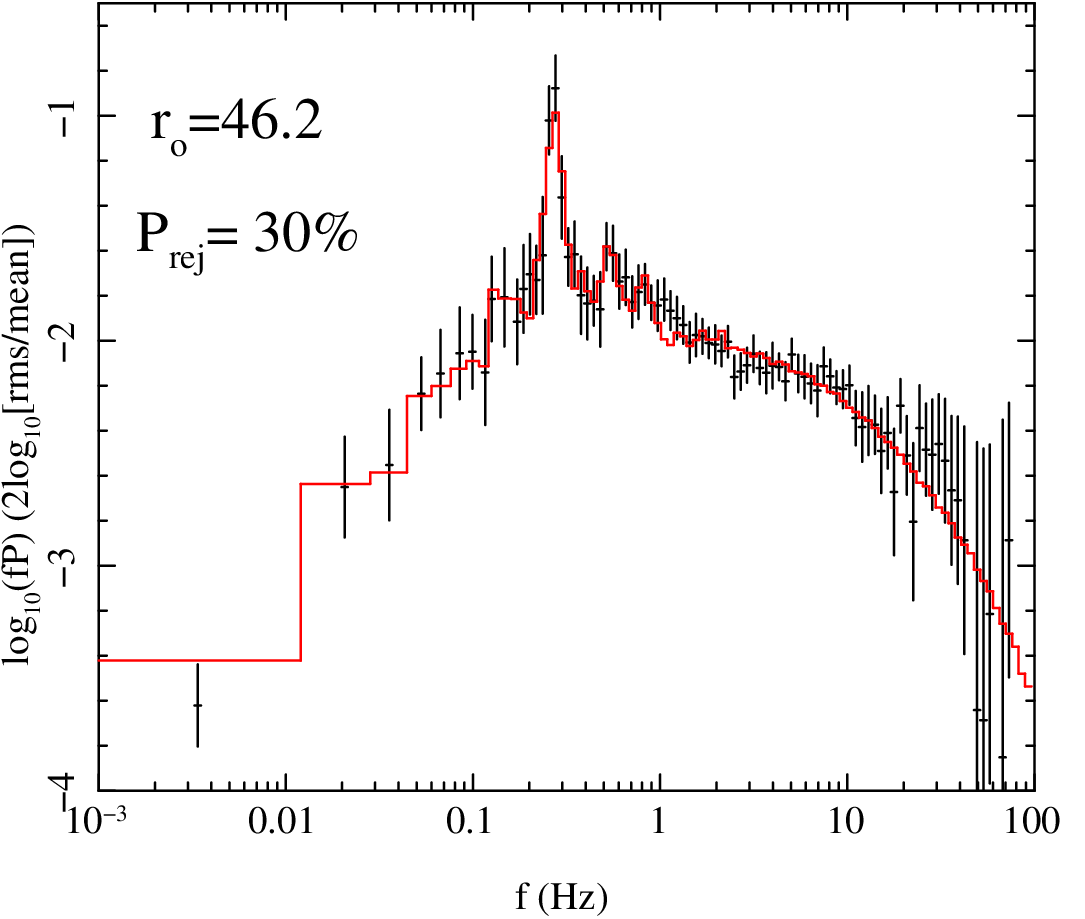}\\
\leavevmode  \epsfxsize=5.5cm \epsfbox{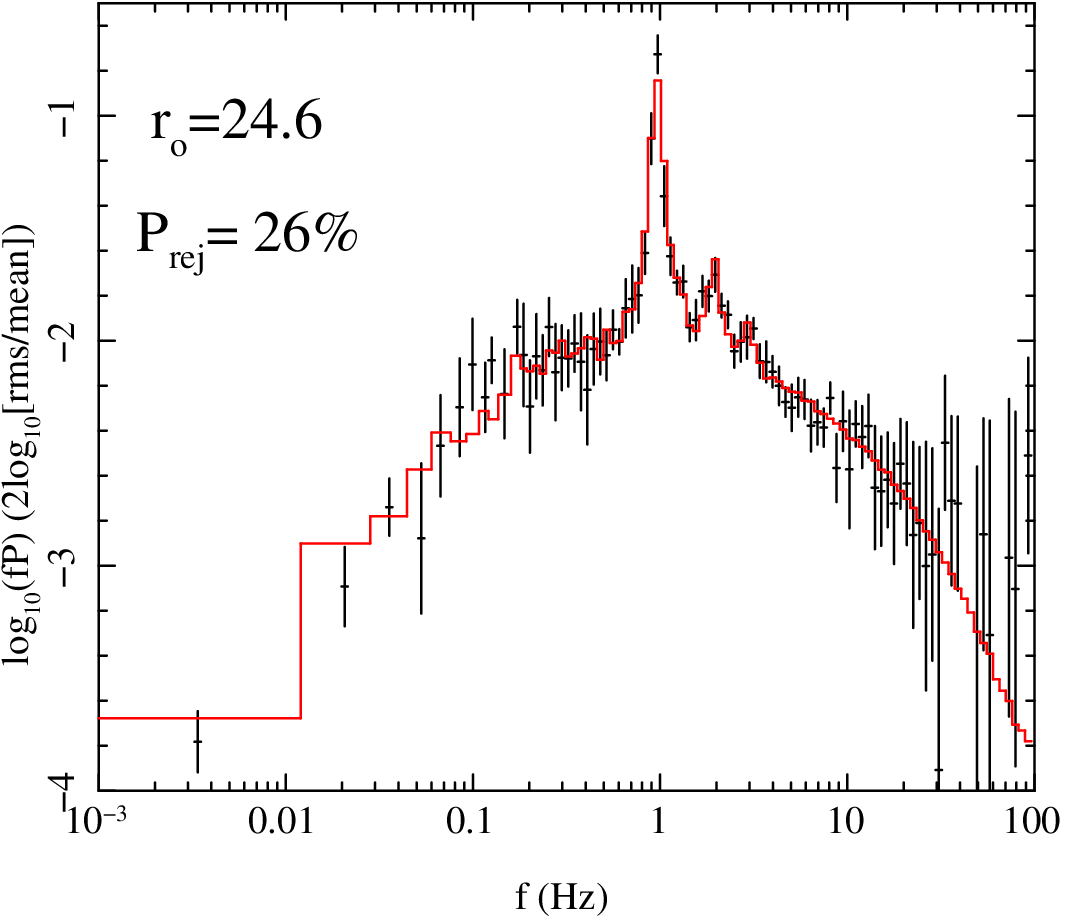}\\
\leavevmode  \epsfxsize=5.5cm \epsfbox{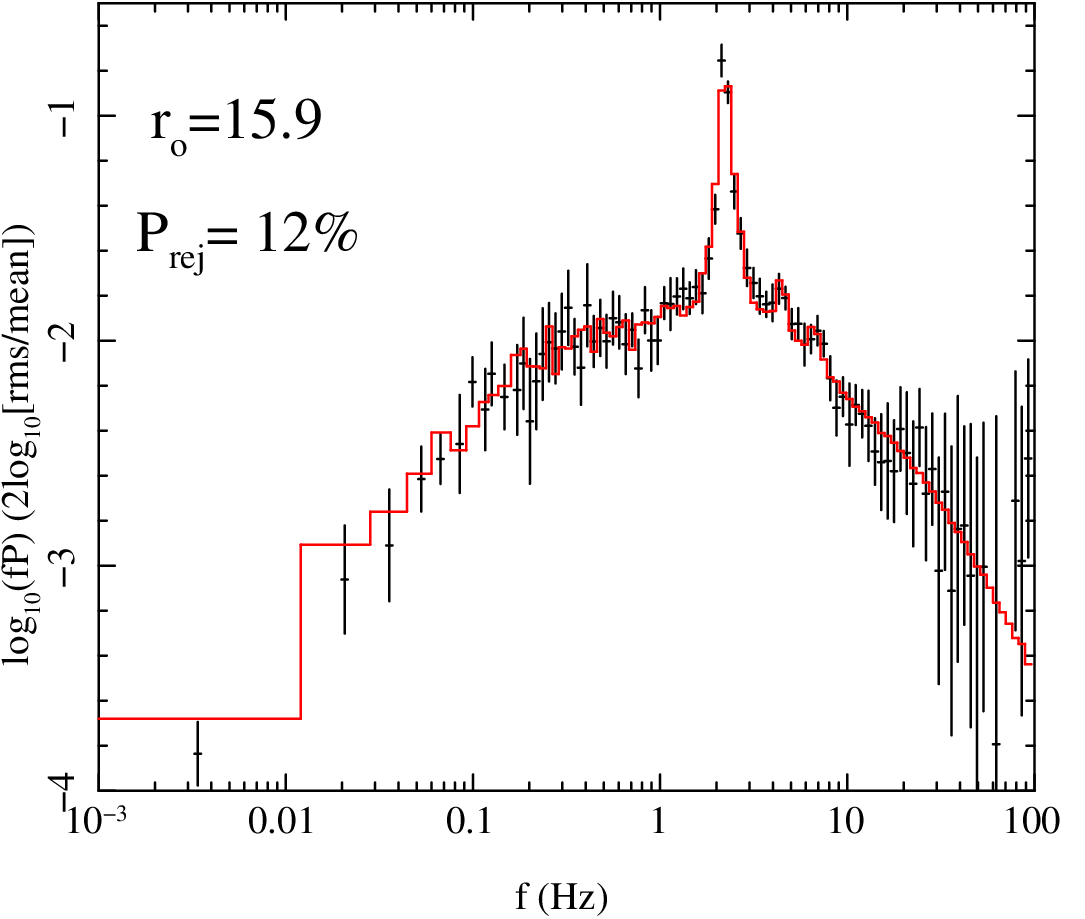}\\
\leavevmode  \epsfxsize=5.5cm \epsfbox{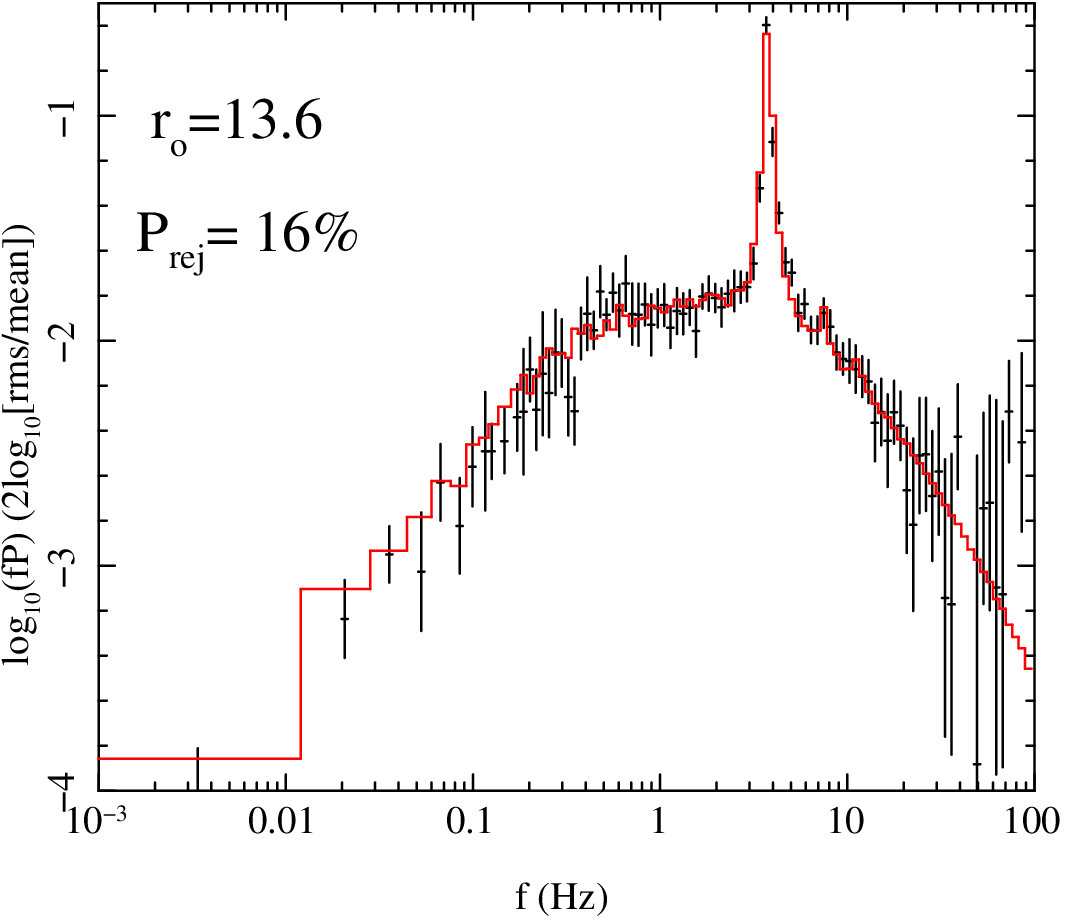}
\end{array}$
\caption{Results of fits for $\gamma=4.15$ and $b(r)=1$ plotted
with the white noise subtracted. The truncation radius, $r_o$, and
rejection probability, $P_{rej}$, are included in the plots.
Here, $r_i=2.45$ and $m=0.46-1.016$.}
\label{fig:r-gammas}
\end{figure}

\begin{figure}
\centering$
\begin{array}{c}
\leavevmode  \epsfxsize=5.5cm \epsfbox{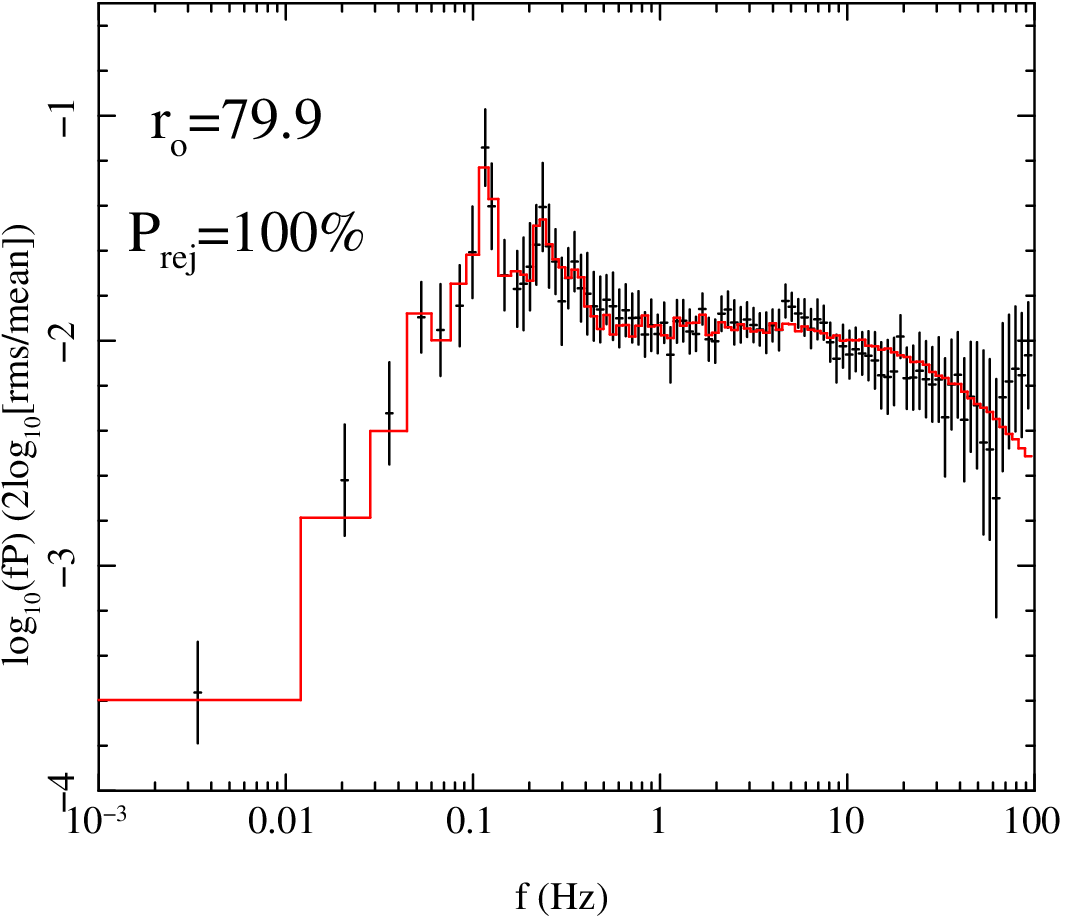}\\
\leavevmode  \epsfxsize=5.5cm \epsfbox{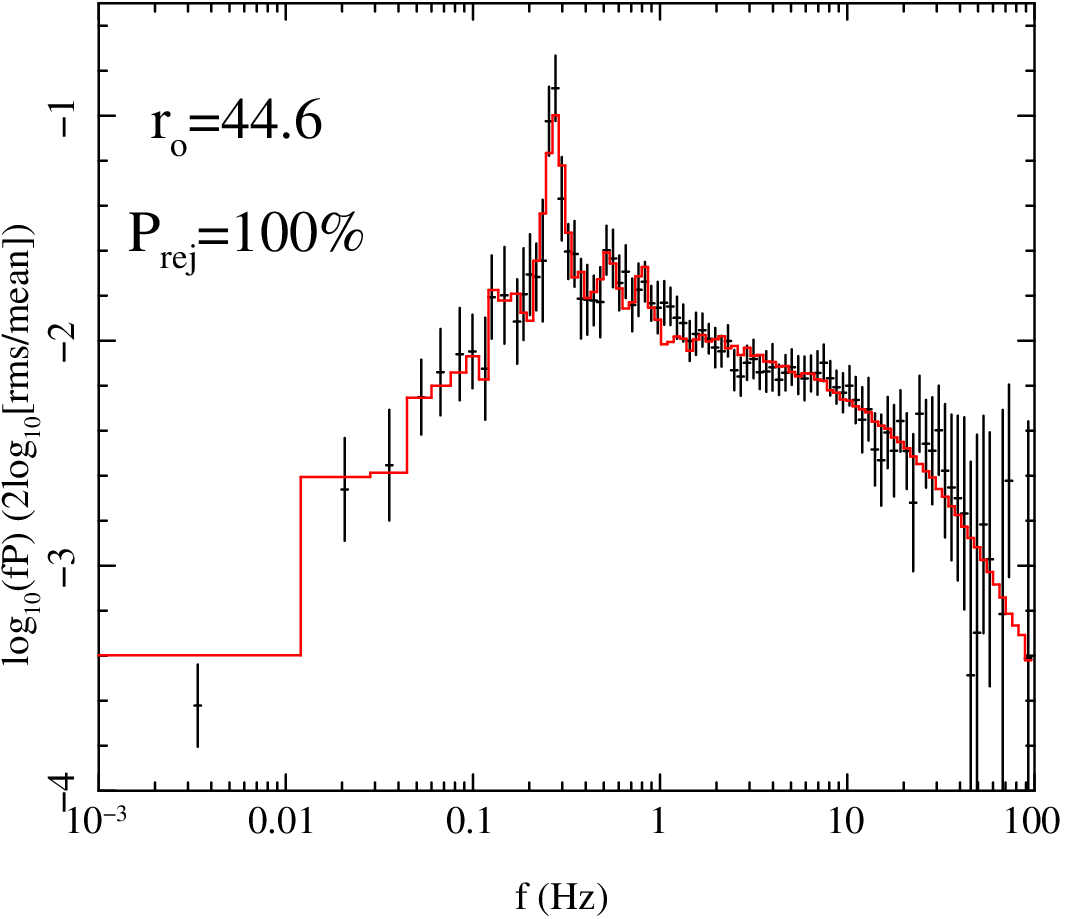}\\
\leavevmode  \epsfxsize=5.5cm \epsfbox{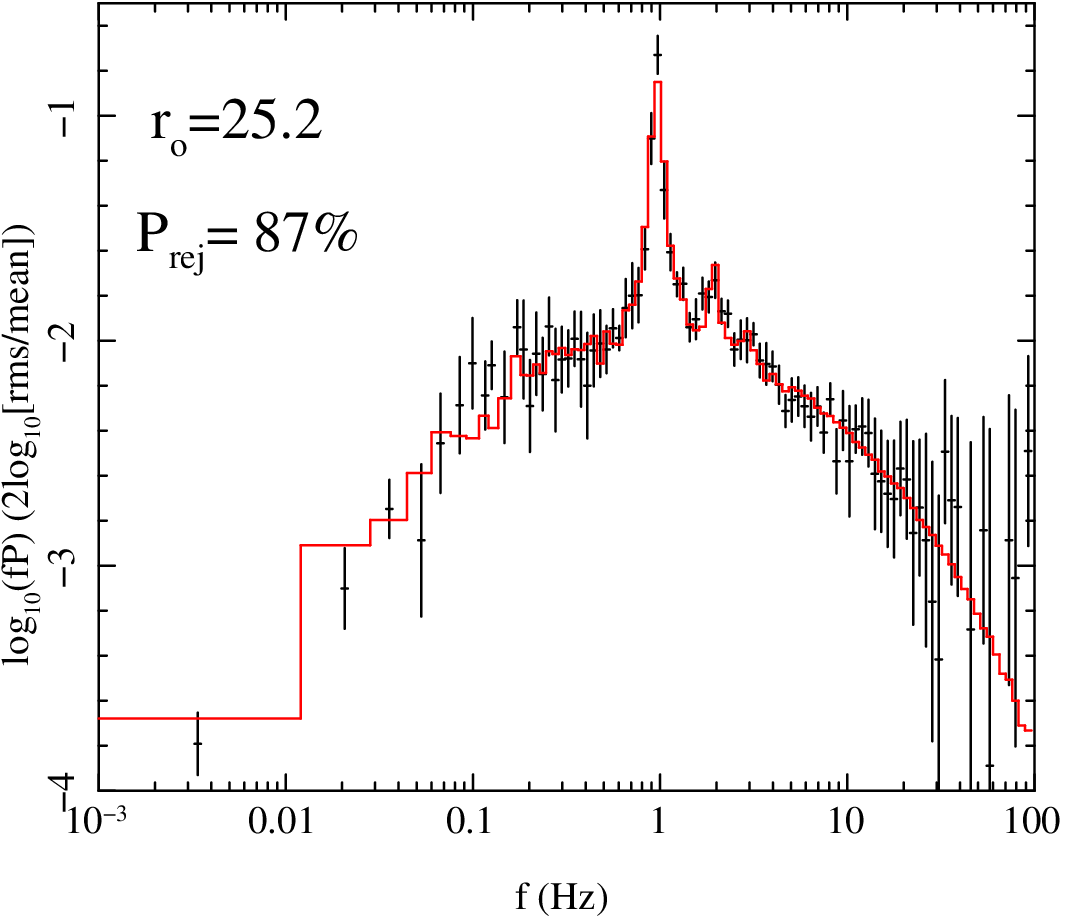}\\
\leavevmode  \epsfxsize=5.5cm \epsfbox{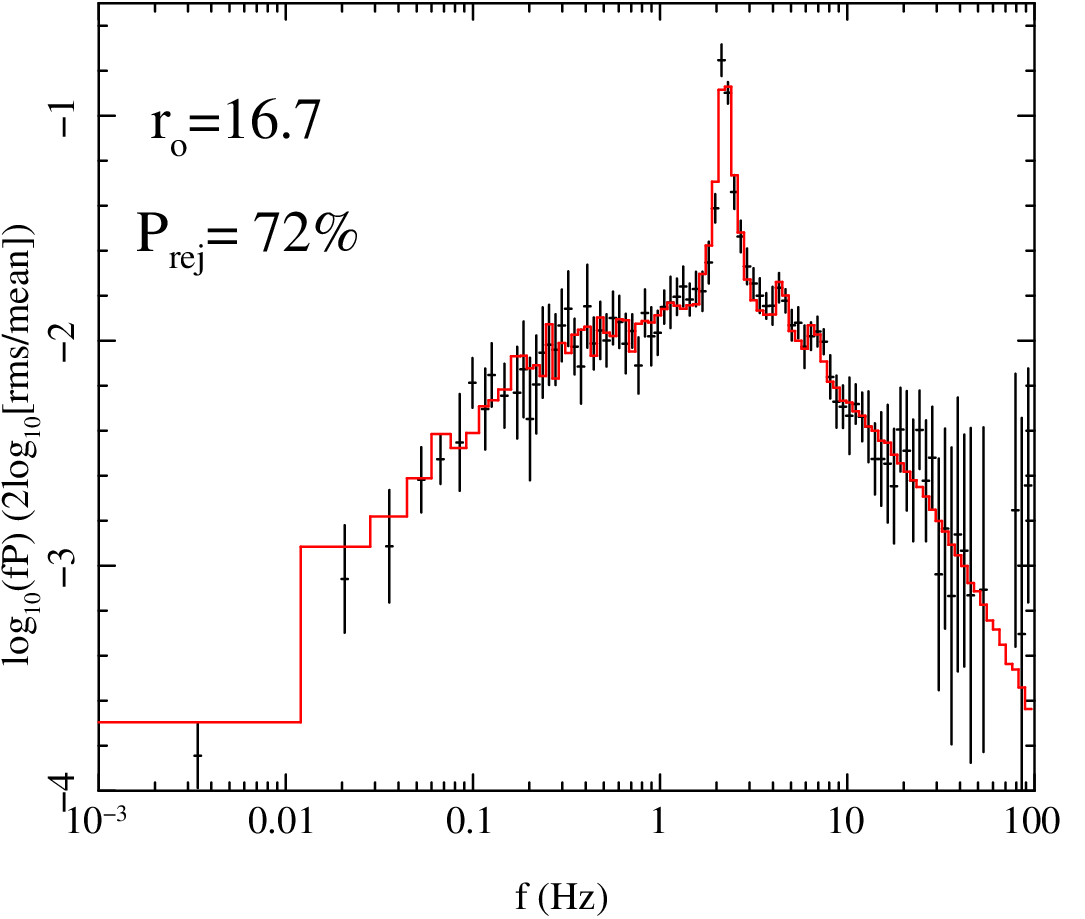}\\
\leavevmode  \epsfxsize=5.5cm \epsfbox{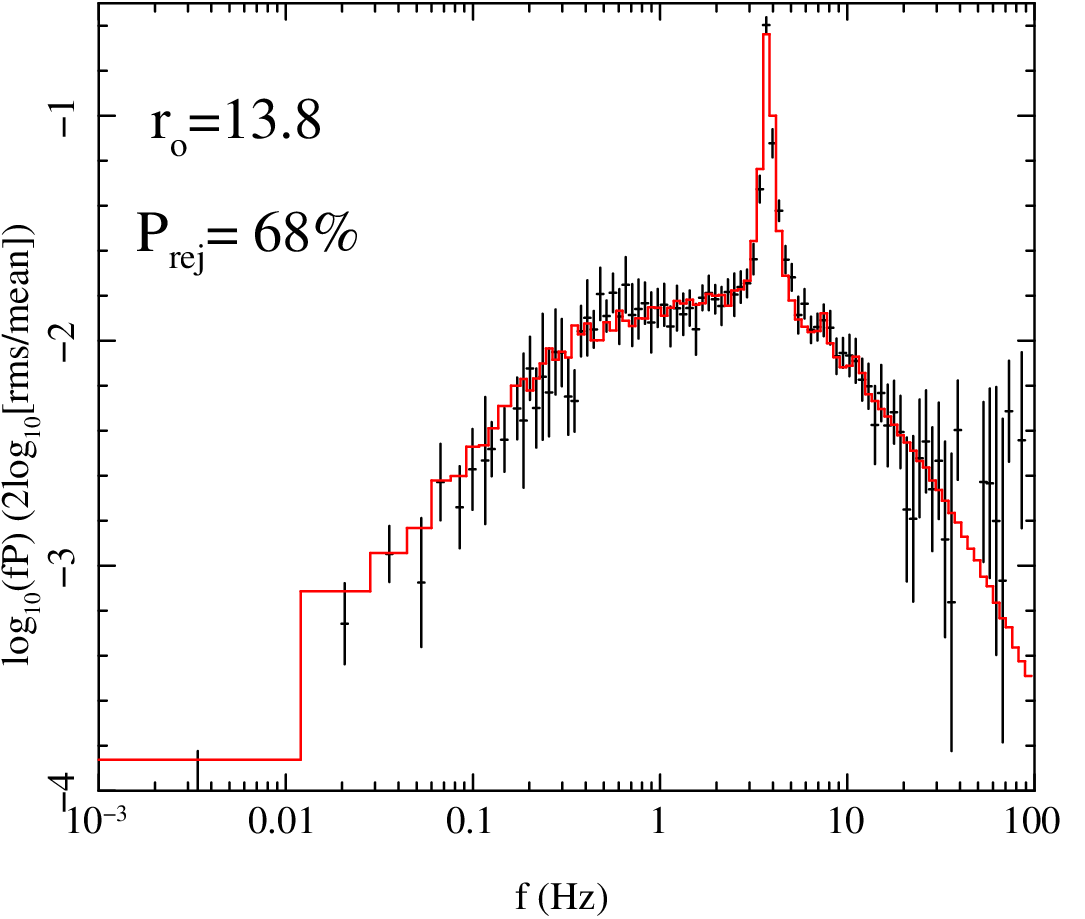}
\end{array}$
\caption{Results of fits for $\gamma=5.06$ and stress free inner boundary
plotted with the white noise subtracted. The truncation radius, $r_o$,
and rejection probability, $P_{rej}$, are included in the plots.
Here, $r_i=2.00$ and $m=0.44-1.30$.}
\label{fig:r-gammasf}
\end{figure}

In this section we fit our model to the $5$ observations considered
assuming a stressed inner boundary condition. Figure \ref{fig:r-gammas}
displays the best fit models and data with the inner radius and emissivity
index required to
be constant across the $5$ observations. These appear by eye to be good
fits,  and this is confirmed by their rejection probabilities of  $51\%$,
$30\%$, $26\%$, $12\%$ and $16\%$ for observations 1--5, respectively. 
We find the inner radius to be $r_i=2.45$ which, although low, is greater
than the horizon of the black hole ($r_H\approx 1.87$ for our chosen $a_*=0.5$).
The truncation radius moves from $r_o=75.7-13.6$. However, the fit requires
that the power law index governing the viscous frequency, $m$, increases across
the observations, from $m=0.46-1.016$. We will discuss the significance of this
in section \ref{sec:para}.

\subsubsection{Stress free inner boundary}
\label{sec:stressfree}

In the previous section, we recovered an inner radius fairly close to the
horizon ($r_i=2.45$) for fits with a stressed boundary condition. This
assumption implies a causal connection to material closer in which, given
the proximity of the emission to the horizon, may not be physical.
Hence the stress free inner boundary condition may be more
physically appropriate. We repeat the fits assuming this, 
with results shown in Figure \ref{fig:r-gammasf}. We recover an even lower
inner radius ($r_i=2.0$), and similar values for the truncation radius
($r_o=79.9-13.8$) and power law index governing the viscous frequency
($m=0.44-1.30$). By eye, these again appear to be excellent fits; in fact
they look nearly identical to Figure \ref{fig:r-gammas} and have only a
marginally higher $\chi^2$ value. However, these have rejection probabilities
of $100\%$, $100\%$, $87\%$, $72\%$ and $68\%$ for observations $1-5$
respectively, so they are all statistically unacceptable. A rejection
probability of $100\%$ means that all of the 512 simulated PSDs had
$\chi^2_{dis}(sim)$ smaller than the one `observed' $\chi^2_{dis}$, making
it extremely unlikely that the observed data is part of the simulated
distribution. 

This discrepancy between $\chi^2$ and rejection probability 
points to the complex nature of the statistics of the periodogram. 
We discuss this in more detail in the Appendix, but here we simply
note that these results do 
\textit{not necessarily} mean that there is no good fit for the
stress free version of the model: they just mean that the real best fit
parameters do not give the lowest $\chi^2$. 
The previous fits with a stressed boundary condition also may not be
the best fit in terms of rejection probability, the difference is that 
with the stressed boundary condition the minimum $\chi^2$ solution
was also a statistically acceptable fit on rejection probability. 
We will explore this in more detail in a future paper (Ingram \& Done, in 
preparation)

However, as the fits appear to be good by eye, it seems
very likely that there should be a set of parameters very close
to the ones that minimize $\chi^2$ that give much lower 
rejection probabilities.

\section{Discussion}
\label{sec:para}

Although the statistical complexities of the
periodogram currently limit the model fitting to data,  the
parameters we derive are still most likely very close to the actual
best fit parameters, and all four fits in the previous section came 
out with similar results for $r_o$ and the viscous frequency radial 
dependence. While the change in $r_o$ is predicted from the
truncated disc/hot inner flow geometry proposed to model the 
correlated spectral changes, the change in viscous frequency 
radial dependence is not. 

There are two (at least) scenarios in which this could occur. Firstly,
this could indicate that the viscous frequency is well approximated by
a power law but that the form of this changes as the truncation radius
moves inwards (as assumed in Ingram \& Done 2010 for the neutron star
power spectra).  There is currently no theoretical model in which this
occurs. A more attractive (to us) possibility is that a power law is a
poor approximation for the viscous frequency and, as the truncation
radius moves, the power law that best approximates the true shape of
the viscous frequency with radius changes. This would imply that there
is a stationary function $f_{visc}(r)$ (and by extension $v_r(r)$)
which would allow this model to fit all 5 observations with only a
change in $r_o$.  We note that exactly this behavior is predicted in
global simulations of an ADAF as the
flow has to accelerate strongly to pass through a sonic point
(e.g. Narayan, Kato \& Honma 1997; Gammie \& Popham 1998). These
models give an in-fall velocity $v_r(r)$ which is well approximated by
a quadratic in log space (see Fig 1 in Gammie \& Popham
1998). This will also allow the model to display a drop off in surface
density in agreement with simulations as discussed in section \ref{sec:qpo}.
Approximating such a velocity law by a power law of index $m$
will result in $m$ increasing as the radial extent of the flow
decreases, similar to the observed trend. We will explicitly fit
such models in a later paper (Ingram \& Done in preparation).

\section{Conclusions}

The truncated disc/hot inner flow model designed to describe the
spectral evolution of BHB can also give a self consistent geometry in
which to model the correlated evolution of the power spectrum.
Propagating fluctuations through a hot flow which extends from and
outer to inner radius, $r_o-r_i$, can produce the band limited noise
characteristic of the continuum power spectrum, as well as producing
the rms-flux relation (L97; K01; AU06).  Lense Thirring precession of
this {\em same} hot flow can produce the QPO, with frequency set by
the {\em same} parameters of $r_o$ and $r_i$, together with the
surface density of the flow (IDF09). The surface density is itself
given self consistently by mass conservation from the propagating
fluctuations.  This predicts that the surface density fluctuates, so
predicts that the QPO frequency will vary on short timescales
(i.e. that it is a quasi rather than true period). These fluctuations
set an upper limit to the coherence of the QPO, and this increases (i.e.
width decreases) as $r_o$ decreases. This is due to the decrease in
fluctuation power due to the smaller range of radii from which to pick
up variability.  All these features are well known properties of the
data (e.g. Remillard \& McClintock 2006; DGK07): this model gives the
first quantitative description of their origin.  The fluctuations also
predict that the flux and QPO frequency are correlated on short
timescales, as a perturbation in the surface density at large radii
leads to a longer QPO frequency but has little effect on the
luminosity. As this propagates down, it weights the mass
distribution to smaller radii, increasing the QPO frequency but also
increasingly contributing to the luminosity due to the centrally peaked
emissivity. This behavior has also recently been
observed (Heil et al 2011).

The model also gives a framework in which to interpret some otherwise
very puzzling aspects of the energy dependence of the
variability seen in BHB. The extended emission region can 
be inhomogeneous, with different parts of the flow producing a
different spectrum. The outermost parts of the flow are closest to the
cool disc, so will intercept more seed photons and have a softer
spectrum than that produced in the more photon-starved inner part of the
flow (Kawabata \& Mineshige 2010; Makishima et al 2008; Takahashi et 
al 2008). This implies that a larger fraction of
the lower energy Compton scattered photons come from larger radii in 
the flow than the higher energy ones. The higher frequency variability
is preferentially produced at the smallest radii, where the spectrum is
hardest. The flow at these small radii is also furthest from the cool
disc, so has little reflection spectrum superimposed on the Compton
continuum. Thus the model predicts that the fastest variability has
the hardest spectrum and smallest reflected fraction, while slower
variability has a softer spectrum and larger reflected fraction. 
This trend is also observed in the data (Revnivtsev et al 1999), and is
very difficult to interpret
in any other geometric picture as the inner disc edge cannot 
change in radius on even the longest timescale (few seconds) 
over which this relation is seen. 

Similarly, the propagating fluctuations model means that a fluctuation
starts at larger radii and then accretes down to smaller radii. Thus the
fluctuation first affects the region producing a softer spectrum, then
propagates down to smaller radii which produce the harder spectrum,
so the hard band lags the soft band. The size of this lag
depends on the frequency of fluctuations considered. Slow fluctuations
(low frequencies) 
are produced at the outermost radius, so have the longest propagation 
time down to the innermost radius. High frequencies are produced 
only close to the inner radius, so only have a short distance to travel and
hence have shorter lags. This gives rise to the frequency dependent
time lags seen in the spectrum (Miyamoto \& Kitamoto 1989; Revnivtsev et al
2001; K01; AU06). In future work,
we will quantitatively test these ideas against real data by incorporating
a radial dependence of the emitted spectrum in our model,
as well as putting in the more complex form for the 
radial dependence of the viscous frequency predicted
from global ADAF models and use improved fit statistics (Ingram \& Done, 
in preparation). We will also calculate the light curve shape predicted
by Lense-Thirring precession of a hot inner flow, where the Comptonised 
emission is modulated by the difference in projected area of the
(translucent) hot flow, self occultation and variation in seed photons
from the different projected area of the disc 
(Ingram, Done \& Zycki in prep). This
predicts non-sinusoidal variability in the light curve i.e. this 
predicts the harmonic structure of the QPO. 

While the many successes of the  model are clearly evident, it is
also clear that it is still far from complete. The most obvious outstanding 
issues are of the interaction of the hot flow with the truncated disc. 
The mechanism by which the cool disc truncates is not well
established, though evaporation powered by thermal conduction
between the two different temperature fluids almost certainly plays
some role in this (Liu et al 1997; Rozanska \& Czerny 2000; Mayer \&
Pringle 2007). Whatever the mechanism, it seems physically
unlikely that this will give a smooth transition between a cool thin
disc and the hot flow. Any inhomogeneities will probably also be
amplified by the difference in velocity between the disc and flow
(discs are close to Keplarian, while the hot flow is strongly sub Keplarian)
so there will be a shearing turbulent layer formed between 
them. Recent results show that there is variability associated with the
truncated disc at a few 10s of seconds in the low/hard state of the
bright BHBs GX339-4 and SWIFT J1753.5-0127 (Wilkinson \& Uttley 2009),
suggesting that there is considerable complexity in the disc truncation
(see also Chiang et al 2010). Full numerical simulations of the MRI
in a composite truncated disc/hot inner flow geometry are probably
required in order to show the effect of these. However, such simulations
are way beyond current computer capabilities. A more tractable issue
is the effect of relativity on the propagating fluctuations.
Near the black hole,
light bending and time dilation should be important and consequently
future versions of this model need to take these effects into
account. The final goal should of course be the creation of a fully
relativistic model which can produce a Fourier resolved spectrum with
both energy and time dependence such that we can test it against
observations such as the PSD, the energy spectrum, the lag spectrum,
the cross spectrum etc. This is of course very ambitious but it is the
only way we can genuinely achieve a full theoretical understanding of
what drives mass accretion and emission in BHBs.

\section{Acknowledgements}

AI acknowledges the support of an STFC studentship.

%-----------------------------------------------------

\appendix

\section{A closer look at the statistics}

Here, we attempt to address some of the statistical problems encountered
in section \ref{sec:data} by focusing in on observation 4 as it is a
particularly striking example. The fit for the stressed version of the
model (hereafter case 1) is plotted in the 4th panel from the top in
Figure \ref{fig:r-gammas} and the fit for the stress free version (hereafter
case 2) is plotted in the 4th panel from top in Figure \ref{fig:r-gammasf}.
By eye, these fits both look excellent and are very difficult to tell apart.
However, the rejection probability is $12\%$ for case 1 and $72\%$ for
case 2! This is extremely confusing, especially considering that case 2
actually has a slightly lower $\chi^2$ ($23.0$ compared to $23.8$) than case
1! We will also discuss the obvious over fitting implied by such low $\chi^2$
values later on in this section.

\begin{figure}
\centering$
\begin{array}{c}
\leavevmode  \epsfxsize=5.5cm \epsfbox{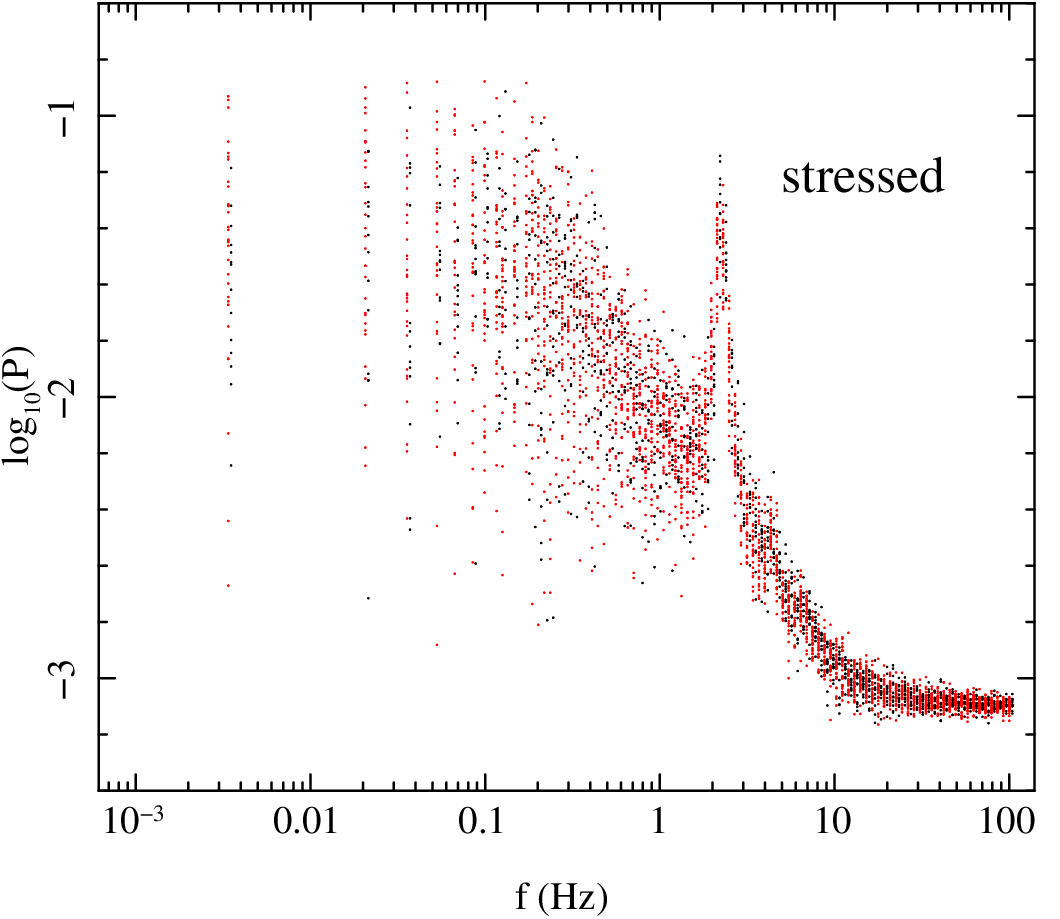}\\
\leavevmode  \epsfxsize=5.5cm \epsfbox{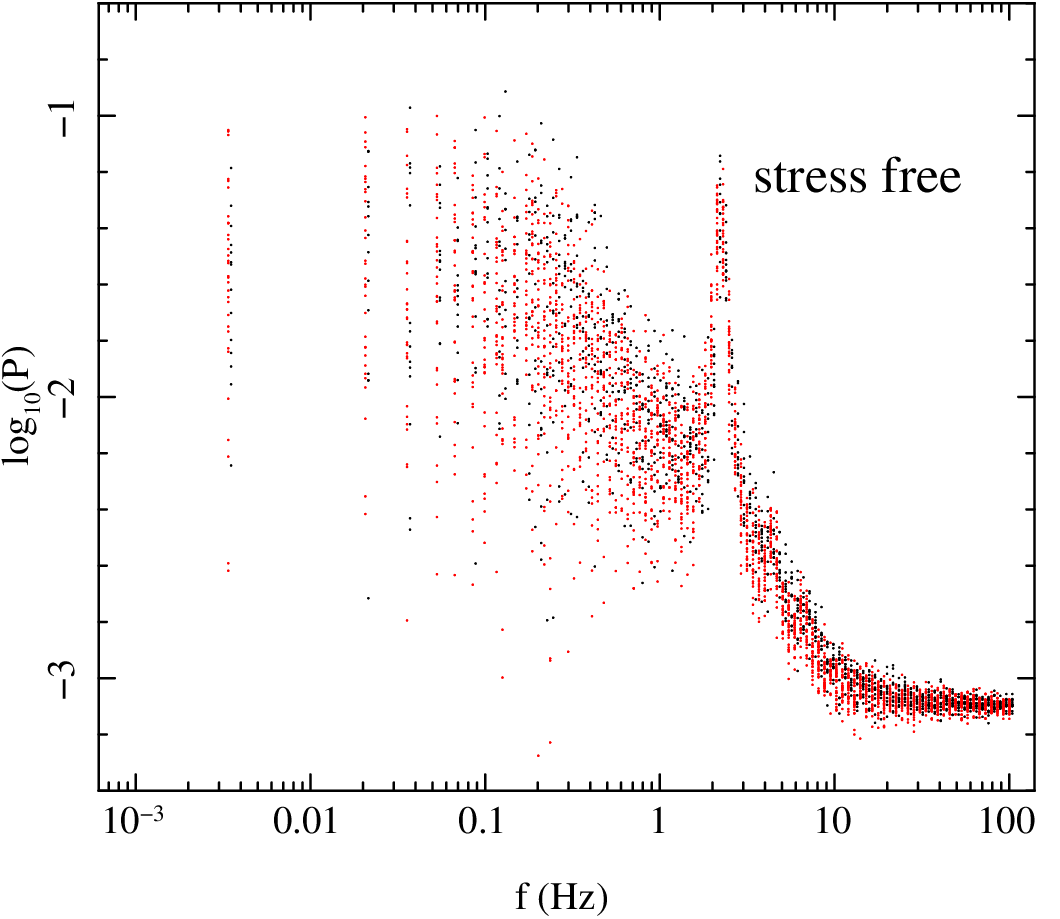}
\end{array}$
\caption{
Raw periodogram points for both data (black) and model (red) with
case 1 on the top and case 2 on the bottom (see text). As this is
the data for observation 4, there are $14$ black points per frequency bin.
Our model calculates smoothed periodograms over $32$ realisations so
there are $32$ red points per frequency bin.
}
\label{fig:scatter}
\end{figure}

\begin{figure}
\centering$
\begin{array}{c}
\leavevmode  \epsfxsize=5.5cm \epsfbox{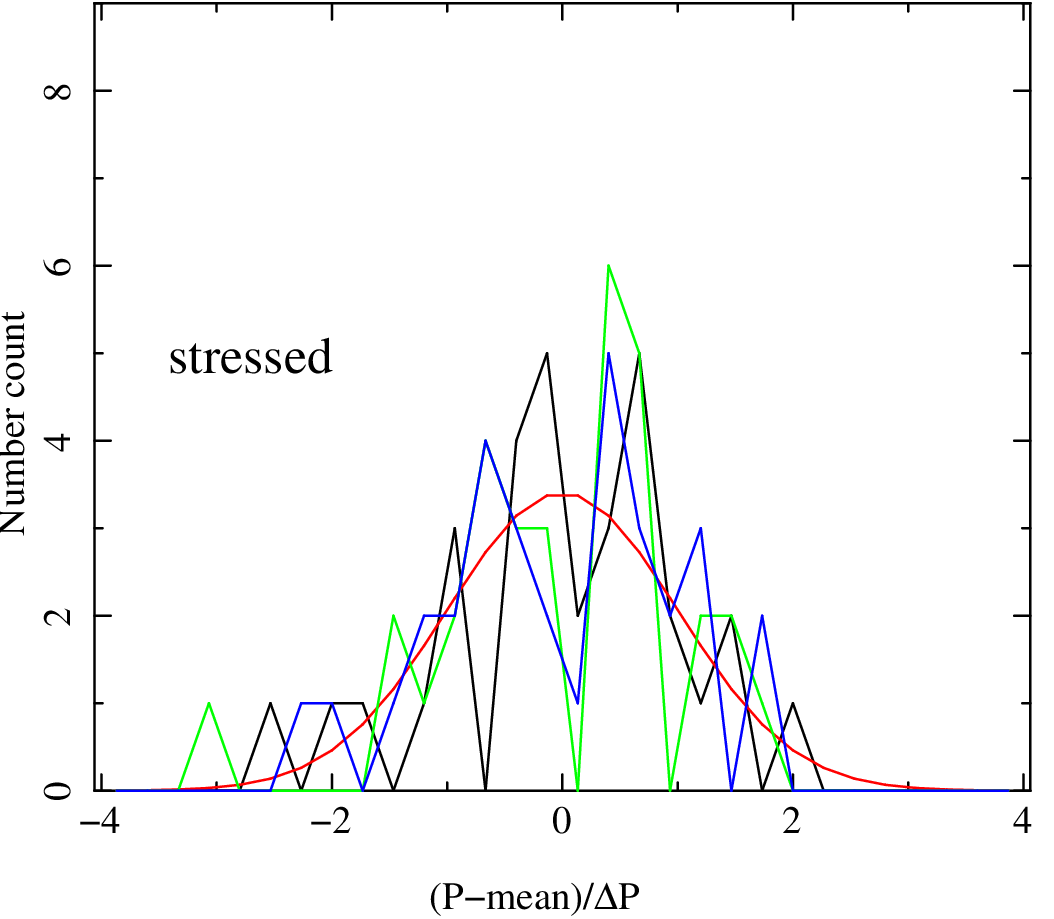}\\
\leavevmode  \epsfxsize=5.5cm \epsfbox{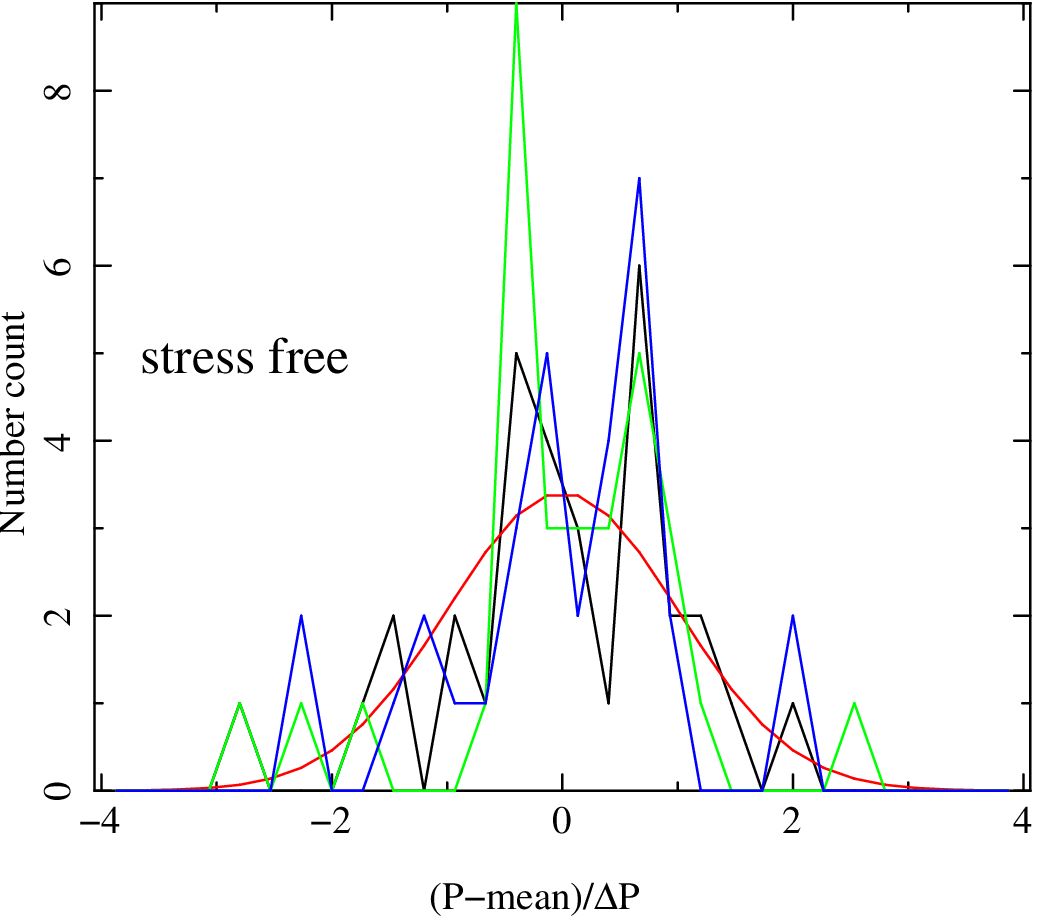}
\end{array}$
\caption{
Histogram showing the distribution of periodogram points at
$f=0.1$Hz (black), $9.5$Hz (green) and $73$Hz (blue) for
case 1 (top) and case 2 (bottom). The three distributions
have been normalised to have the same mean for clarity. We
can see that these distributions have departures from a Gaussian
(red line), especially in case 2.
}
\label{fig:dis2}
\end{figure}

In an attempt to understand this discrepancy, we look a little closer at these
two examples. It turns out that $\chi^2_{dis}$ for case 2 is larger than in
case 1 ($59207$ for case 2; $53074$ for case 1). Although this sheds a
little light on the mystery, it doesn't explain such a huge descrepancy in
rejection propbability. However, the $\chi^2_{dis}(sim)$ values are all much
smaller for case 2 meaning that the simulated values agree with each other much
better than they do with the observed values, making it extremely unlikely that
the observed points lie on the distribution of simulated points. In an attempt
to see this difference, we plot all of the raw periodogram points for both cases
in Figure \ref{fig:scatter}. The black and red points are observed and simulated
respectively with case 1 on the top and case 2 on the bottom. We see that the two
plots are indistinguishable.

In Figure \ref{fig:dis2}, we have looked at 3 specific frequencies: $0.1$Hz (black),
$9.5$Hz (green) and $73$Hz (blue) and created a histogram to see the distribution of
periodogram values. Note, the three distributions have been normalised to have
a mean of $0$ and a standard deviation of $1$ for clarity. Again, case 1 is on the
top. Although there are departures from Gaussian behaviour in all of these
distributions, it is clear that the distributions for case 2 are far more centrally
peaked than those of case 1. This is consistent with $\chi^2_{dis}(sim)$ being
smaller in general for case 2. There are also hints of baises which may account
for the apparent over fitting implied by such small $\chi^2$ values.

In our analysis, we calculated the logarithmically smoothed periodogram over
$M=32$ realisations in order to calculate and minimise $\chi^2$. We did this
because Papadakis \& Lawrence (1993) show that the logarithmically smoothed
periodogram can have a Gaussian distribution if it is smoothed over more than
$M=20$ realisations, in comparision to the smoothed periodogram (i.e. no
logarithms) where you would need more than $M=50$ realisations. This implies that our
logarithmically smoothed periodograms have a Gaussian error distribution,
therefore making $\chi^2$ an appropriate fit statistic. However, there are two
problems with this, both arising because the error distribution depends on the
underlying power spectrum. Papadakis \& Lawrence (1993) were looking at red
noise which is a significantly different variability process than that considered
here. Therefore, just because 20 realisations are enough for them, doesn't mean
it is enough for us. More significantly, the same authors show that the
logarithmically smoothed periodogram may give a baised estimate of a power spectrum
with a high curvature. Since, we are simulating QPOs, we are in this very regeim
which could explain the hint of bais displayed in Figure \ref{fig:dis2}.

For these reasons, in a future paper (Ingram \& Done in prep) we will use
smoothed periodograms (no logarithms) and attempt to find a way around the
resulting non-Gaussian errors. To do this, we will consider statistical methods
more general than $\chi^2$ but less computationally intensive than minimising
the rejection probability. Hopefully this will allows us to overcome these
statistical problems.

\label{lastpage}

\end{document}